\documentclass[longauth]{aa} 
\usepackage{graphicx}
\usepackage{txfonts}
\usepackage{natbib}
\bibpunct{(}{)}{;}{a}{}{,}
\newcommand{\msun}{{M$_\odot$}}
\begin{document}

\title{The Vimos VLT Deep Survey:}
\subtitle{\bf Global properties of 20000 galaxies in the $\mathrm{I_{AB}}<22.5$
WIDE survey
}

\author{
B. Garilli \inst{1}
\and O. Le F\`evre \inst{2}
\and L. Guzzo \inst{9}
\and D. Maccagni \inst{1}
\and V. Le Brun \inst{2}
\and S. de la Torre \inst{2}
\and B. Meneux \inst{1,9}
\and L. Tresse \inst{2}
\and P. Franzetti \inst{1}
\and G. Zamorani \inst{3} 
\and A. Zanichelli \inst{4}
\and L. Gregorini \inst{4}
\and D. Vergani \inst{1}
\and D. Bottini \inst{1}
\and R. Scaramella \inst{4,13}
\and M. Scodeggio \inst{1}
\and G. Vettolani \inst{4}
\and C. Adami \inst{2}
\and S. Arnouts \inst{22,2}
\and S. Bardelli  \inst{3}
\and M. Bolzonella  \inst{3} 
\and A. Cappi    \inst{3}
\and S. Charlot \inst{8,10}
\and P. Ciliegi    \inst{3}  
\and T. Contini \inst{7}
\and S. Foucaud \inst{21}
\and I. Gavignaud \inst{12}
\and O. Ilbert \inst{20}
\and A. Iovino \inst{9}
\and F. Lamareille \inst{7}
\and H.J. McCracken \inst{10,11}
\and B. Marano     \inst{6}  
\and C. Marinoni \inst{18}
\and A. Mazure \inst{2}
\and R. Merighi   \inst{3} 
\and S. Paltani \inst{15,16}
\and R. Pell\`o \inst{7}
\and A. Pollo \inst{2,17}
\and L. Pozzetti    \inst{3} 
\and M. Radovich \inst{5}
\and E. Zucca    \inst{3}
\and Blaizot, J.\inst{23}
\and A. Bongiorno \inst{21}
\and O. Cucciati \inst{9,14}
\and Y. Mellier \inst{10,11}
\and C. Moreau \inst{2}
\and L. Paioro \inst{1}
}

\institute{
INAF-IASF, Via Bassini 15, I-20133, Milano, Italy
\and
 Laboratoire d'Astrophysique de Marseille, UMR 6110 CNRS-Universit\'e
 de Provence, BP8, F-13376 Marseille Cedex 12, France 
\and
INAF-Osservatorio Astronomico di Bologna, Via Ranzani 1, I-40127,
Bologna, Italy
\and
INAF-IRA, Via Gobetti 101, I-40129, Bologna, Italy 
\and
INAF-Osservatorio Astronomico di Capodimonte, Via Moiariello 16,
I-80131, Napoli, Italy
Italy
\and
Universit\`a di Bologna, Dipartimento di Astronomia, Via Ranzani 1,
I-40127, Bologna, Italy
\and
Laboratoire d' Astrophysique de Toulouse-Tarbes, Universit\'e de Toulouse,
CNRS, 14 avenue Edouard Belin, F-31400 Toulouse, France
\and
Max Planck Institut f\"ur Astrophysik, D-85741, Garching, Germany
\and
INAF-Osservatorio Astronomico di Brera, Via Brera 28, I-20021, Milan, Italy 
Italy
\and
Institut d'Astrophysique de Paris, UMR 7095, 98 bis Bvd Arago, F-75014, Paris, France 
\and 
Observatoire de Paris, LERMA, 61 Avenue de l'Observatoire, F-75014, Paris, France 
\and 
Astrophysical Institute Potsdam, An der Sternwarte 16, D-14482, Potsdam, Germany 
\and 
INAF-Osservatorio Astronomico di Roma, Via di Frascati 33, I-00040, Monte Porzio Catone, Italy 
\and 
Universit\'a di Milano-Bicocca, Dipartimento di Fisica, Piazza delle Scienze 3, I-20126, Milano, Italy
\and
Integral Science Data Centre, ch. d'\'Ecogia 16, CH-1290, Versoix, Switzerland 
\and
Geneva Observatory, ch. des Maillettes 51, CH-1290, Sauverny, Switzerland 
\and
Astronomical Observatory of the Jagiellonian University, ul Orla 171, PL-30-244, Krak{\'o}w, Poland 
\and 
Centre de Physique Th\'eorique, UMR 6207 CNRS-Universit\'e de Provence, F-13288, Marseille, France 
%19
%\and 
%Centro de Astrof{\'{i}}sica da Universidade do Porto, Rua das Estrelas, P-4150-762, Porto, Portugal 
\and
Institute for Astronomy, 2680 Woodlawn Dr., University of Hawaii, Honolulu, Hawaii, 96822, USA 
\and
School of Physics \& Astronomy, University of Nottingham,
University Park, Nottingham, NG72RD, UK
\and
Max Planck Institut f\"ur Extraterrestrische Physik (MPE), Giessenbachstrasse 1,
D-85748 Garching bei M\"unchen,Germany
\and
Canada France Hawaii Telescope corporation, Mamalahoa Hwy,  
Kamuela, HI-96743, USA
\and
Universit\'e de Lyon, Lyon, F-69003, France ; Universit\'e Lyon 1,
Observatoire de Lyon, 9 avenue Charles Andr\'e, Saint-Genis Laval, F-69230,
France ; CNRS, UMR 5574, Centre de Recherche Astrophysique de Lyon ; Ecole
Normale Sup\'erieure de Lyon, Lyon, F-69007, France.
}

\offprints{B.Garilli, bianca@lambrate.inaf.it}

\date{Received ...; accepted ...}

\abstract 
{The VVDS-Wide survey has been designed with the general aim of tracing
the large-scale distribution of galaxies at
$z\sim 1$
on comoving scales reaching 
$\sim100 h^{-1}~Mpc$, while providing a good control of cosmic variance
over areas
as large as a few square degrees.  This is achieved by measuring
redshifts with VIMOS at the ESO VLT to a limiting magnitude
$I_{AB}=22.5$, targeting four 
independent fields with size up to 4 $deg^{2}$ each.} 
{We discuss here the survey
strategy which covers 8.6  $deg^{2}$ and present the general properties of the current
redshift sample.  This includes 32734 spectra in the four regions, covering a
total area of 6.1 $deg^{2}$ with a sampling rate of 22 to 24\%.  This
paper accompanies the public release of the first 18143 redshifts of
the VVDS-Wide survey from the 4 deg$^2$ contiguous area of the F22
field at RA=22$^h$.}
{We have devised and tested an objective method to assess the quality
of each spectrum, providing a compact figure-of-merit,
particularly effective in the case of long-lasting 
spectroscopic surveys with varying observing conditions.
Our figure
of merit  
is a measure of 
the 
robustness of the redshift measurement and, most importantly,
can be used to select galaxies with uniform high-quality spectra 
to carry out reliable measurements of spectral features. We use the data
available over the four independent regions to directly measure the
variance in galaxy counts.  We compare it with general predictions
from the observed galaxy two-point correlation function at different
redshifts and with that measured in mock galaxy surveys built from the
Millennium simulation.}
{{\bf The purely magnitude-limited VVDS Wide
sample includes 19977 galaxies, 304 type I AGNs, and 9913 stars. 
The redshift success rate 
is above 90\% 
independently of magnitude.
A cone diagram of the galaxy spatial distribution provides us with the
current largest overview of large-scale structure up to z$\sim 1$, 
showing  
a rich texture of over- and under-dense regions.
We give the mean N(z) distribution averaged over 6.1 
$\deg^{2}$ for a sample limited in magnitude to
$\mathrm{I_{AB}}=22.5$.
Comparing galaxy densities from the four fields  
shows that in a redshift bin $\Delta$$z= 0.1$ at $z\sim1$ one
still has factor-of-two variations over areas as large as $\sim0.25$
deg$^{2}$.  This level of cosmic variance agrees with that obtained
by integrating the galaxy two-point correlation function estimated from
the F22 field alone.  It is also in fairly good statistical agreement
with that predicted by the Millennium mocks.}}
{{\bf The VVDS WIDE survey provides the currently largest area coverage
  among redshift surveys reaching z$\sim 1$.  The variance estimated
  over the survey fields shows explicitly how clustering
  results from deep surveys of even ~1 deg$^2$ size should be
  interpreted with caution.  The survey data represent a rich
    data base to
    select complete sub-samples of high-quality spectra and to study
    galaxy ensemble properties and galaxy clustering over
    unprecedented scales at these redshifts.
The redshift catalog of the 
4 deg$^{2}$ F22 field is publicly available at http://cencosw.oamp.fr.}}


\authorrunning{B.Garilli et al.}
\titlerunning{VVDS - The VVDS Wide sample}

\maketitle

\section{Introduction}
The large-scale 
distribution of galaxies contains unique information on the structure of 
our Universe and the fundamental parameters of the cosmological model.
The relation of galaxy properties to large-scale structure in turn
provides important clues on the physics of galaxy formation within the
standard paradigm in which baryons are assembled inside dark-matter
halos \citep[e.g.][]{white_rees}.  Redshift surveys of the ``local''
($z<0.2$) Universe as the 2dFGRS \citep{2df} and SDSS \citep{sdss}
contain several hundred thousand 
galaxies spanning a few thousands square degrees.
These large samples and explored volumes 
have allowed large-scale structure studies to be pushed well into the linear
regime $r\gg 5$ h$^{-1}$ Mpc 
while 
having at the same time a detailed characterization of small-scale
clustering and its dependence on galaxy properties like luminosity, colour and
morphology  \citep[e.g.][]{2df_clus_type,2df_clus1,2df_clus2,sdss_clus1,sdss_clus2}.
All these features and properties are expected to depend on redshift,
and different evolutionary paths can lead to similar observational
properties in the local universe. 
Ideally, one would 
like to be able to gather similarly large samples over comparably large 
volumes, at cosmologically relevant distances ($z>>0.3$). 
First pioneering deep redshift surveys capable of
measuring the evolution of clustering go back to the 1990's and were
limited to a few hundred square arcminutes
\citep[e.g.][]{cfrs_clustering,cnoc_clustering}.
Even deeper measurements of clustering evolution were provided by
specific color-selected surveys, using the Lyman-break technique,
although these give a very biased view of large-scale structure
limited to a specific class of objects \citep[e.g.][]{steidel98}.
More recent surveys like GOODS \citep[e.g.][]{goods} and DEEP
\citep[e.g.][]{deep}
provide extended
multi-wavelength coverage, but are still limited to small fields.
Only recently, thanks to the increased
multi-plexing ability of spectrographs mounted on 10-m class telescopes, 
robust clustering studies of the general galaxy population at $z\sim
1$ have become feasible. This opportunity has been exploited by the
VVDS \citep{vvds_main} and the DEEP2 \citep{deep2} surveys.
The VVDS Deep sample \citep[][]{vvds_main}, in particular, covered a
reasonably large area ($\sim 0.5$ $deg^{2}$)  up to redshift 4 and 
to a very deep
magnitude limit ($I_{AB}=24$).  Major clustering results using these
data have included studies of the evolution of galaxy 
clustering since $z\sim 2$ \citep[][]{vvds_clus}, 
its dependence on luminosity, 
spectral type and stellar mass \citep{vvds_clus_lum,vvds_clus_type,vvds_clus_mass}
and the evolution and non-linearity of 
galaxy bias \citep[][]{vvds_bias}, together with a direct assessment
of the evolution of enviromental effects, as the dependence of colour
\citep{vvds_environment} or luminosity function
\citep{vvds_lf,vvds_lf_type,vvds_lf_environment} on local density. 
Still, the area surveyed
by the VVDS Deep is not yet large enough to fully
characterize large-scale structure at high redshift: results from 2dF
show that structures of size $\sim50h^{-1}$ Mpc do exist in the local
Universe, while in the VVDS-Deep itself a structure at 
$z\sim0.9$ is found to fill the full survey field  ($\sim$14
h$^{-1}$ Mpc) \citep{vvds_nature} The {\it Wide} part of the VVDS survey has been conceived specifically to
improve upon this, covering structures with size $\sim 50$ h$^{-1}$
Mpc at $z\sim 1$, while having the ability to measure the variance in
galaxy density on scales of a few square degrees.  This will be achieved by
measuring redshifts to $I_{AB}=22.5$ over four separated fields on
the sky, with size up to 4 deg$^2$ each.\\
In this paper we present a first analysis of the currently available
redshifts from the VVDS-Wide spectroscopic survey, including in particular
the data collected over the full $\sim 4$ 
deg$^2$ area of the F22 field, which are publicly released to the
scientific community.  The paper is organized as follows.
In section 2 we describe the VVDS Wide survey strategy and report on
the status of the observations conducted so far; in section 3 we 
assess redshift reliability depending on data quality,
in section 4 we present the main
characteristics of the resulting redshift catalog, while
in section 5 we present the widest cone diagram currently available up to 
$z \sim 1.0$, quantify the field to field variance of the redshift 
ditribution and how it can affect smaller size surveys and compare the
observed cosmic variance with model predictions.
\\
Throughout this paper, we have used a Concordance Cosmology with
$\Omega_{\rm m}=0.3$, and $\Omega_{\Lambda}=0.7$. 
The Hubble constant is normally parameterized via $\mathrm{h=H}_{0}/100$, while
a value $\mathrm{H}_{0}=70~\mathrm{km~s^{-1}~Mpc^{-1}}$
has been used when computing absolute magnitudes. 
\\
\\
\section{The VVDS Wide survey}
\begin{figure}
\resizebox{\hsize}{!}{\includegraphics[clip=true]{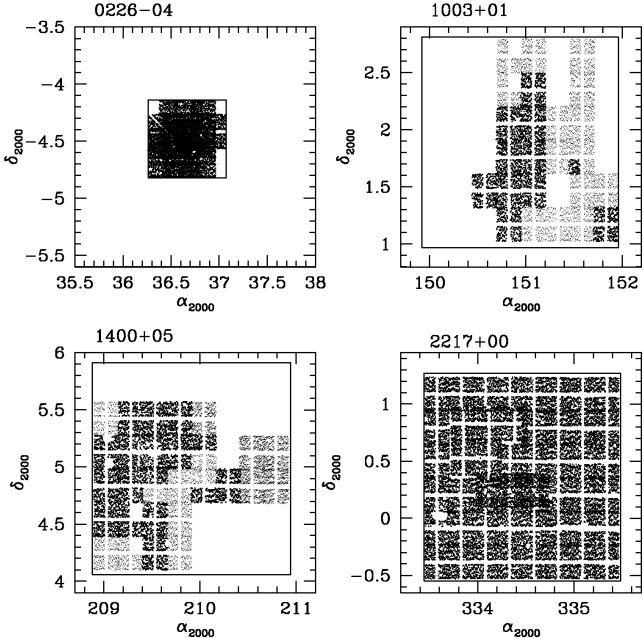}}
\caption{Layout of observed fields for the VVDS Wide survey: the
  square represents 
  the planned area to be covered. 
  Black dots correspond to 
  measured redshifts which are used in this paper, while grey dots are
  from objects which have been observed, but whose redshift is still
  being finalised. The empty grid corresponds to the VIMOS
  foot-print, which leaves a 2-arcmin-thick empty cross among the four
  quadrants.   
  All data for the F02 and F22 fields are public at http://cencosw.oamp.fr/.  }
\label{coverage}
\end{figure}

The VVDS Wide survey uses VIMOS at the ESO VLT to target 
4 separate fields,  
one of which includes the VVDS Deep survey area,
evenly distributed on the sky and covering a total of
16 $\mathrm{deg}^{2}$.
With a $2\times 2$ deg$^2$ size, each field
can span along the diagonal a transverse comoving size of 116 h$^{-1}$
Mpc at $z=1$. 
The names and coordinates of each field are given in
Table~\ref{summaryTable}.\\
In each of the areas we have excellent photometric coverage,
extending from U to K. In addition to the U, ${\it BVRI}$, ${\it JK}$
surveys conducted by the VVDS team 
(\cite{vvds_imaging_u}; \cite{vvds_imaging_f02};
\cite{vvds_imaging_jk}), the sky regions at 02 and 22 hours are also
covered by the CFHTLS survey\footnote{
Based on observations obtained with MegaPrime/MegaCam, a joint  
project of CFHT and CEA/DAPNIA, at the Canada-France-Hawaii Telescope  
(CFHT) which is operated by the National Research Council (NRC) of  
Canada, the Institut National des Science de l'Univers of the Centre  
National de la Recherche Scientifique (CNRS) of France, and the  
University of Hawaii. This work is based in part on data products  
produced at TERAPIX and the Canadian Astronomy Data Centre as part of  
the Canada-France-Hawaii Telescope Legacy Survey, a collaborative  
project of NRC and CNRS.
}
and the UKIDSS survey \citep{UKIDSS}.
The VVDS Deep field has also 
been observed at 1.4 GHz at the VLA
(\cite{vvds_radio};\cite{vvds_radio_2}), by XMM (\cite{vvds_LSS};\cite{vvds_XMDS}), by Galex
(\cite{vvds_imaging_uv}; \cite{vvds_galex_lumden}) and by Spitzer
\citep{vvds_swire}. 
The spectroscopic sample has been derived 
from an I selected
photometric catalog applying a pure flux limit at 
$\mathrm{I_{AB}}=22.5$, which provides the best compromise between
efficiency in covering a large area and depth of the final
spectroscopic sample.    
A specific choice of the survey was that of not removing stars {\it
  a priori}  using colour or compactness 
criteria, to avoid biases against compact galaxies and AGN.\\
The original plan of the VVDS Wide survey, 
involved a ``two-pass'' observing strategy:
each of the four areas is covered by two slightly displaced (2
  arcmin) grids of adjacent VIMOS pointings 
(see below for a 
description of the current
implementation of this plan). 
This strategy allows one
  to reach a spectroscopic sampling rate of $\sim 35\%$ of all
  galaxies with $\mathrm{I_{AB}}<22.5$, which is important for density
  reconstruction studies \citep{vvds_bias}, 
while keeping the required observing time 
within reasonable limits. At the same time, the
  2-arcmin shifts is chosen as   
to fill (at least partially)
the gaps left by the VIMOS footprint. 
For the VVDS Wide survey, the exposure time of each pointing was 
45 minutes
in MOS mode, using the Low Resolution Red grism. 
As in the case of the VVDS Deep survey \citep[see][]{vvds_main}, we have used a
jitter observing sequence, with 5 steps along the slit, each
separated by 0.7 arcsec. This strategy allows us to considerably
reduce the fringing produced by the CCDs above 
$\sim 8000 \AA$ \citep{vvds_tech},
although fringing residuals still appear for the brighter and more
extended sources, as well as in those observations where seeing was
higher than 1.0 arcsec. 
Observation preparation, mask
layout, and observing strategy is the same as for the VVDS Deep
sample: using the VMMPS software 
\citep{vmmps}, we have been able to place $\sim 400$ slits on average
per 
VIMOS mask-set
down to the limiting magnitude of the VVDS Wide survey.
Data have been reduced using the VIMOS Interactive Pipeline and
Graphical Interface package \citep[VIPGI][]{vipgi}. \\
The observations presented here were collected mostly during
Guaranteed Time observations (5 extended visitor observing runs
from Oct. 2002 to Sep. 2004), with a small fraction acquired during two
further runs in Guest Observer standard time (service runs in February
2006 and 2007).   
As visually summarized in Fig.~\ref{coverage},
we completed the first pass on the
4.0 $\mathrm{deg}^{2}$ F22 area, plus a second pass on the central
$\sim 0.5\times 0.5$ deg$^2$ of the same field. We further include
here the redshift measurements from the first pass over 0.8 and 1.2
$\mathrm{deg}^{2}$ in F10 and F14 respectively, while
further 2.1 deg$^2$ in these area are under analysis and are not
included in this paper (grey dots in Fig.~\ref{coverage}). 
Finally,
we also include redshifts for all 
galaxies with $I_{AB}<22.5$ in the $0.5\ deg^{2}$ of the F02 field
covered by the {\it VVDS-Deep} survey to $I_{AB}=24$.  Overall, this
data set corresponds to $\sim $36\% of the original VVDS Wide survey goal.\\
Given the instrument geometry, with one pass only 
there are {\it empty crosses} not covered
by the instrument (see Fig.~\ref{coverage}, F10, F14 and outer part of
F22 areas). In Table ~\ref{summaryTable} we give both the total
area covered so far by the VVDS Wide survey, (i.e. the global area covered
by either grey or black points in fig.~\ref{coverage}), and the {\it
  effective} area, i.e. the area including only fully reduced pointings (black
points only) and net of the {\it
  empty crosses}.
In Table ~\ref{summaryTable} we also give the total number of
pointings reduced so far for each field, and the average sampling rate
of measured redshifts at the given magnitude limit.\\ 
%data summary table
\begin{table*}
\caption{VVDS Wide survey field position and observing information}
\label{summaryTable}
\centering
\begin{tabular}{c c c c c c c}
\hline\hline
Field & R.A & Dec & Surveyed area & Effective area & N.of pointings& sampling rate  \\
\hline
0226-04 (F02) & 02h26m00.0s & -04deg30'00'' &  0.5 &  0.5  & 20~~ & 24\%\\
1003+01 (F10) & 10h03m00.0s & +01deg30'00'' &  1.9 &  0.6  & 11$^1$ & 24\%\\
1400+05 (F14) & 14h00m00.0s & +05deg00'00'' &  2.2 &  0.9  & 17$^2$ & 22\%\\
2217+00 (F22) & 22h17m50.4s & +00deg24'00'' &  4.0 &  3.0  & 51~~ & 22\%\\
\hline
\end{tabular}
\begin{tabular}{l c}
$^1$For 1 pointing, only 1 quadrant has been reduced so far
&~~~~~~~~~~~~~~~~~~~~~~~~~~~~~~~~~~~~~~~~~~~~~~~~~~~~~~~~~~~~~~~~~~~~~~~~~~~~~~~~~~~~~~~~~~~~~~~~~~~~~~\\
$^2$Reduction of 4 pointings is still partial&\\
\end{tabular}
\end{table*}

%area coverage figure
% macro coverage.mac comando c
\section{Redshift measurement, data quality, and reliability}
Redshifts have been measured  using the same ``double-check'' procedure
described in \citet{vvds_main}, adopting the same
grading scheme to characterize the reliability of the measured redshift: 
\begin{itemize} 
\item
    flag 4: a 100\% secure redshift, with high SNR spectrum and
    obvious spectral features supporting the redshift measurement;  
\item
    flag 3: a very secure redshift, strong spectral features; 
\item
    flag 2: a secure redshift measurement, several features in support
    of the measurement; 
\item
    flag 1: a tentative redshift measurement, based on weak spectral features
    and continuum shape;  
\item
    flag 0: no redshift measurement possible, no apparent features; 
\item
    flag 9: only one secure single spectral feature in emission,
    tipically interpreted as [OII]3727 Å, or $\rm H\alpha$.
\end{itemize}
A similar classification is used for broad line AGN, which we
identify as spectra showing at least one "broad line'' (i.e. resolved
at the spectral resolution of the VVDS). Flags for broad line AGN have the
following meaning 
\begin{itemize}
\item
    flag 14: secure AGN with 100\% secure redshift, at least
    2 broad lines; 
\item
    flag 13: secure AGN with good confidence redshift, based on
    one broad line and some faint additional feature; 
\item
    flag 19: secure AGN with one single secure emission line feature,
    redshift based on one line only;
\item
    flag 12: a 100\% secure redshift measurement, but lines are not
    significantly broad, might not be an AGN;
\item
    flag 11: a tentative redshift measurement, spectral features not
    significantly broad. 
\end{itemize}
Serendipitous objects
appearing by chance within the slit of the main target are identified by
adding a ``2'' in front of the flag.
We have classified with flag = -10 objects in slits with a clear
observational 
problem, like e.g. objects for which the automated spectra
extraction algorithm in VIPGI \citep[][]{vipgi} failed, 
or objects too close to the edge
of a slit to allow for a proper sky subtraction.
In the following, redshifts with a flag between 2 and 9 (or 12 and 19
in the case of AGN) are referred to as secure redshifts.
\subsection{Data quality}
When conducting a large spectroscopic survey, carried out over
years, under different weather conditions, and with
different people involved at different times in the data reduction and 
redshift measurement process, it is important to identify an objective way
to assess the quality of the data and of the reduction process,
independent on the redshift measurement success or failure.
For the VVDS Deep and Wide surveys, we have devised a method which
takes into account the most important 
observational/reduction factors.\\
Slit obscuration due to field vignetting, effective exposure time, seeing and sky
transparency directly impact on the number of photons collected for each
spectrum; sky brightness at constant exposure time determines the S/N ratio,
and the quality of the wavelength calibration has an impact on the accuracy of
the redshift measurement. 
The goal we set was to devise an objective quality parameter which could be
used to make an a priori selection of the best data at hand. 
The final figure of merit we assigned to each spectrum is the
combination of all these factors in such a way that the highest is the
figure of merit, the highest the spectrum quality. 
In the following, we shall
discuss in turn each 
separate 
contribution to this quality parameter, show the overall
results for both surveys and relate it to the redshift confidence
level.
\subsubsection{Slit obscuration}
\label{obscured_slits}
VIMOS takes advantage of the full Nasmyth
field of view, but, due to the design of the
guiding probe, a fraction of
the field of view can be partially vignetted for some positions of the guiding star. This has happened especially during
the first observations, before enough experience had been gained on
the choice of the guiding star. 
Obscured slits can be easily identified by looking at the average level
of sky counts in each slit, and comparing it with the average
sky level for all the slits in the quadrant. When the single slit
has a sky level which is lower than 70\% the average sky level, 
the slit has been flagged as ``bad'' assigning to it 
an $obscur_\mathrm{flag}=0$. This happens for a 
total of 75 objects in the VVDS Deep data, and 181 in the VVDS Wide data.
An {\it a posteriori} check shows that these slits account for 68\% and 50\%
of the
spectra where no object is detected in the VVDS Deep
and VVDS Wide sample respectively.
As all the contributions to the quality parameter are eventually 
combined in a multiplicative way, all
obscured slits will end up in a global quality parameter equal to zero.
\subsubsection{Wavelength calibration}
\label{Wavelength calibration}
\begin{figure}
\resizebox{\hsize}{!}{\includegraphics[clip=true]{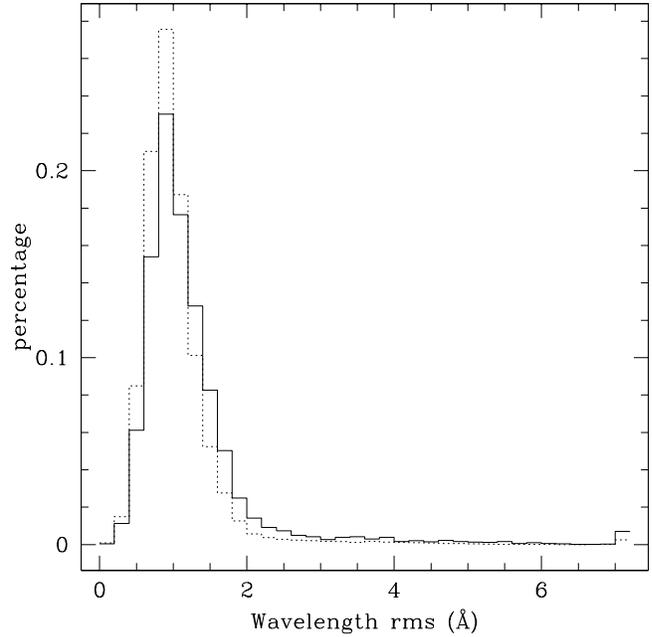}}
\caption{Distribution of the wavelength calibration rms: VVDS Deep survey,
  solid histogram, VVDS Wide survey, dotted histogram
}
\label{lambdahisto}
\end{figure}

As described in \citet{vipgi}, wavelength calibration is performed
using both a global fitting, and a slit per slit
refinement. For some particular slits towards the edge
of the field of view, and in particularly unfavourable positions of the
instrument during the observation, flexures can be important, and it is not possible to get 
a wavelength calibration of the same quality as usual.
Using the wavelength calibration {\it rms} for each slit \citep[see][]{vipgi} as
a measure of the wavelength calibration quality, we can identify
such deviant cases. The distribution of the wavelength calibration {\it rms}
for the VVDS Deep
(solid line) and the VVDS Wide (dotted line) survey spectra
is 
shown in Fig.~\ref{lambdahisto}. There is a small number of slits (1\% both in the VVDS Deep
survey, and in the VVDS Wide survey) showing 
a wavelength calibration rms above $6\AA$ , which essentially means that
wavelength calibration has totally failed.
Such slits get a "wavelength
calibration quality flag" 
$\lambda_\mathrm{flag} =0$. 
% lambdahisto
% macro ~/test_vimos/test_sn/results/lambda_qual.mac comando l
Furthermore, for some pointings the arc calibration exposure was not usable, and
we were forced to calibrate the corresponding spectra using the sky
lines. In such 
cases, the wavelength calibration is never as good as in the standard
case,
since the sky lines are often broad and/or unresolved at our resolution. This
is reflected by the distribution of the wavelength calibration {\it
  rms} for that specific quadrant/pointing, which peaks at $\sim
3.5\AA$ rather than the usual $1.2\AA$. 
We have considered these slits as an intermediate category, and
assigned to them a wavelength calibration quality flag of -1. 
As all the contributions to the quality parameter are eventually 
combined in a multiplicative way, a shaky
( {\it rms} $> 3\AA$) or bad ( {\it rms} $> 6\AA$) wavelength calibration will end
up in a global quality parameter being negative or equal to zero respectively.
\subsubsection{Sky brightness}
Sky brightness depends (at zero order) on moon phase and moon distance.
In principle, knowing these two parameters and using
some calibration table, the expected sky brightness could be
computed. In practice, as
we are interested in 
the global background level, a simpler approach has been adopted: 
\begin{enumerate}
\item
for each slit, the mean one dimensional sky  
spectrum is obtained by taking the median over the sky
 two-dimensional image along the spatial direction, to reject border effects;
\item
after having discarded obscured slits (see \ref{obscured_slits}), all one
dimensional sky spectra
 are combined and the median sky spectrum
 for the whole quadrant derived;
\item
such median sky spectrum is then normalized for the
 exposure time and integrated over the full
wavelength range to get the median sky value
 for that quadrant in that pointing  ($med_\mathrm{sky}$);
\item
by comparing the median sky values obtained for the same quadrant in
 the different pointings, we define a ``reference sky value''
 ($ref_\mathrm{sky}$) as 
the mean of the three lowest "median sky values";
\item
finally, the median sky value per quadrant per pointing is
 compared to $ref_\mathrm{sky}$ for that quadrant, 
and a sky quality factor is defined as\\
$sky_\mathrm{qual}=\frac{1.}{\sqrt{\frac{med_\mathrm{sky}}{ref_\mathrm{sky}}}}$\\
\end{enumerate}
In Fig.~\ref{qualityPar}, top left, the distribution of the sky
quality parameter for the VVDS Deep (solid line) and VVDS Wide (dotted line)
surveys is shown. Overall, the VVDS Wide Survey shows a sky
quality parameter distribution broader than the VVDS Deep survey. This is 
expected, as the VVDS Deep survey observations have been carried
out during dark time, while the VVDS Wide survey ones have been
partially performed during grey time.
\subsubsection{Exposure Time}
The exposure time for each pointing of the VVDS Deep survey had been
planned to be 16200 seconds, while for the Wide it should have been
2700 seconds per pointing. As a matter of fact, in some cases the effective
exposure time has been less than what foreseen, mainly because
metereological conditions had badly deteriorated during the observation,
and of course exposure time has a direct impact on 
the signal to noise ratio as a multiplicative factor. 
There are also a
few observations, performed during visitor runs, which have been
lengthened in the attempt to compensate for high airmass or unstable
metereological conditions.
By comparing the actual total exposure time of
each quadrant in each pointing ($obs_\mathrm{time}$), to the nominal exposure time ``a
priori'' established for the survey ($ref_\mathrm{time}$), we can
define\\

$time_\mathrm{qual}=\sqrt{\frac{obs_\mathrm{time}}{ref_\mathrm{time}}}$\\

In Fig.~\ref{qualityPar}, top right, the distribution of the Exposure Time
quality parameter for the VVDS Deep (solid line) and VVDS Wide (dotted line)
surveys is shown. For the vast majority of the pointings, the exposure
time effectively used is what had been foreseen for that depth.
\subsubsection{Sky Transparency, seeing and slit losses}
Atmospheric conditions have a direct influence on sky 
transparency and seeing, 
which in turn, and coupled with slit losses,
contribute to flux losses in a way which is
not possible to disentangle. In order to estimate their global
contribution, 
an empirical approach has been adopted: 
for each object, we have integrated its spectrum under the I filter
response curve, and compared the thus obtained
$Iflux_\mathrm{spectro}$ to the equivalent quantity as obtained 
from photometry ($Iflux_\mathrm{phot}$).
Then one could in 
principle compute in one shot the effect of seeing, slit losses and transparency on
S/N as\\

$mag_\mathrm{qual}=\sqrt{\frac{Iflux_\mathrm{spectro}}{Iflux_\mathrm{phot}}}$\\

Such ratio should always be below one, by definition, but, as shown in
\citet{vvds_main}, 
a small fraction (around few percent) of objects have a value of
$mag_\mathrm{qual}$ 
above 1.0. This is
due to a number of second-order effects affecting the measure, such as: 1) 
$Iflux_\mathrm{spectro}$ is affected by how well
zero orders or fringing residuals have been removed; 2)
$Iflux_\mathrm{phot}$
has its own errors, larger for fainter magnitudes (~0.2 mag for
objects fainter than $\mathrm{I_{AB}} \sim 23$, for the VVDS Deep survey,
\citet{vvds_imaging_f02});
3) the brighter and more 
extended the object, the more inaccurate is the sky subtraction:
the sky region that can be used to compute the sky level is
small and dominated by pixels affected by slit edge effects. This can
lead to an underestimate of the sky level.
This more often occurs in the VVDS Wide survey pointings, where
the fraction of brighter ($\sim$ larger) objects is higher, and/or in bad
seeing conditions.
To quantify the overall contribution of such second
order effects, we can define \\

$\frac{Iflux_\mathrm{spectro}}{Iflux_\mathrm{phot}}=transmission +
residuals$\\

where $transmission$ is
actually due to sky transparency and seeing/slit width ratio, 
and, for pointlike sources, 
should be constant within one observation.
$residuals$ represent the
contribution of all the second order effects like those listed above, 
and as such can vary from object to object. 
In optimal atmospheric conditions, we should have
$transmission\sim1$ 
and  $residuals=0$. Indeed,
in the magnitude range between $\mathrm{I_{AB}} \sim 21.5$ and
$\mathrm{I_{AB}} \sim 22.5$, where the error on photometric magnitude is negligible and
object sizes are small enough not to be affected by slit losses, or to hamper a good background
estimate, the mean flux ratio is
always below one, being affected by sky transparency only. 
Thus, on a per quadrant and per pointing basis,
using only the range  $21.5<=I_{AB}<=22.5$
we can compute the mean transmission as\\

$transmission=<\frac{Iflux_\mathrm{spectro}}{Iflux_\mathrm{phot}}>$\\

Subsequently, and for each object $i$ for which
$Iflux_\mathrm{spectro}(i)/Iflux_\mathrm{phot}(i)$ is above one, we can estimate
the residuals as \\

$residuals(i)=\frac{Iflux_\mathrm{spectro}(i)}{Iflux_\mathrm{phot}(i)} -
transmission$\\

Finally, the contribution of all these factors to the observation quality
can be computed as\\

$mag_\mathrm{qual}(i)=\sqrt{transmission}*\sqrt{1-residuals(i)}$\\

In Fig.~\ref{qualityPar}, panel {\it e}, the distribution of the $mag_\mathrm{qual}$ parameter for the VVDS Deep (solid line) and VVDS Wide (dashed line)
surveys is shown. Also objects for which no redshift has been
measured are included in this panel. Panel {\it c}
and {\it d} show the distribution of the same parameter
for stars and galaxies separately. 
Comparing the two distributions
obtained for stars, we see that in the VVDS Deep survey most stars
have a figure of merit close to 1, while for the VVDS Wide survey the
peak is around 0.8. As stars are less affected by slit losses than
galaxies, panel {\it c} tells us that the overall better figure of
merit for the VVDS Deep survey is mainly due to the better average
atmospheric conditions during observations.
\subsubsection{Global quality parameter}
The above quality parameters have been computed for each object and 
combined in a multiplicative way as \\

$qual=obscur_\mathrm{flag}*\lambda_\mathrm{flag}*sky_\mathrm{qual}*time_\mathrm{qual}*mag_\mathrm{qual}$\\

% quality parameter factors
% ~/test_vimos/test_sn/results/check_qual_star_gal.mac, comando wide_all
\begin{figure}
\resizebox{\hsize}{!}{\includegraphics[clip=true]{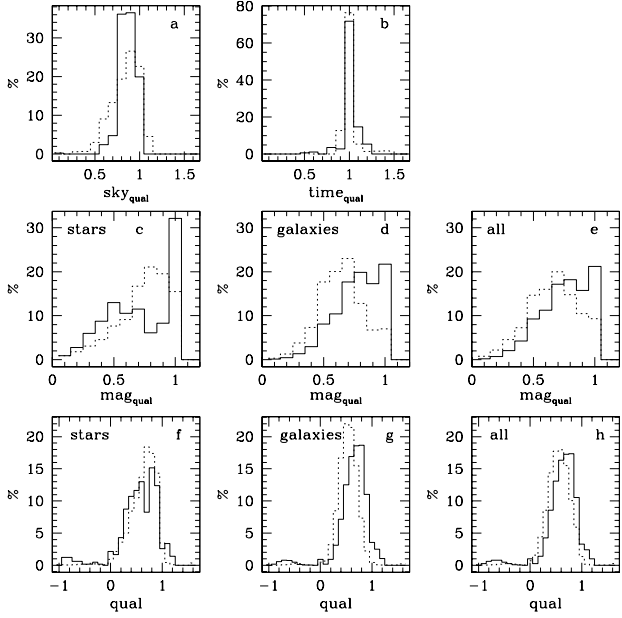}}
\caption{Distribution of the different contributions to the quality
  parameter and of the resulting figure of merit for the VVDS Deep (solid line)
  and VVDS Wide (dashed line) surveys. Panel (a): $sky_\mathrm{qual}$;
(b) $time_\mathrm{qual}$; (c) $mag_\mathrm{qual}$ for spectroscopic
  stars; (d) $mag_\mathrm{qual}$ for galaxies; (e) $mag_\mathrm{qual}$
for the total sample (including failed redshift measurements); (f)
  figure of merit for stars and (g) for galaxies; (h) figure of merit
  for the whole sample (including failed redshift measurements).
}
\label{qualityPar}
\end{figure}

The resulting distribution is shown in Fig.~\ref{qualityPar} bottom panels, for the VVDS Deep (solid
line) and VVDS Wide (dotted line) data.
Negative values of quality pertain to 
objects with poor wavelength calibration, while a quality parameter of
0 is due to either bad wavelength calibration or to obscured slits.
From Fig.~\ref{qualityPar},
panel {\it h}, it is apparent that the global distribution of
the quality parameter is (slightly) better for the  VVDS Deep survey data than
for the VVDS Wide survey data. This is essentially
due to the $mag_\mathrm{qual}$ parameter, which is better for the deep survey,
while the VVDS Wide survey, in which large and extended galaxies 
are more abundant, is globally more affected by slit
loss effects.\\
The percentage of galaxies with quality above 0.5 is 79\% in
the VVDS Deep sample and 58\% in the VVDS Wide sample.
We note, however, that the quality parameter is not an absolute
measure of data quality, but just a relative one: it allows 
to select the best quality spectra we have in our samples (i.e. those
for which slit losses are small, observed for the nominal exposure time
in excellent atmospheric conditions), or conversely, to discard
those data for which something during observations or reduction went
wrong. We will see in the following section that this does not necessarily
prevent, nor assures, 100\% reliable redshift measurements.
\subsubsection{Data quality parameter and redshift flag}
Once the global quality parameter is obtained, it is interesting to
see how it relates to the redshift flag.
If both estimates are reliable, we expect that 
objects with a bad value of the 
quality parameter (i.e. below 0.5) should have a higher probability of
an unsuccessful redshift measurement (flag =0),  
while objects with a good value of the quality parameter (above 0.5) 
should have a higher probability of a
very secure redshift  flag (i.e. 3 or 4). Still, we do not expect 
the opposite to be
totally true, i.e. there may exist spectra with a not so good quality
but to which the redshift can be securely assigned: in fact, the quality
parameter is related to the continuum intensity, its signal to
noise, and
the absolute flux calibration of the data, while the 
flag is a measure of the reliability of the redshift, and is
strongly affected by the presence, or absence, of prominent 
emission/absorption lines.\\
% quality parameter versus flag
% macro ~/test_vimos/test_sn/results/qhisto.mac comando qd
% qdeep.dat (qwide.dat)
% where file qwide/deep.dat come from topcat via collating
% quality_all.dat with Fxx_spec.dat
\begin{figure}
\resizebox{\hsize}{!}{\includegraphics[clip=true]{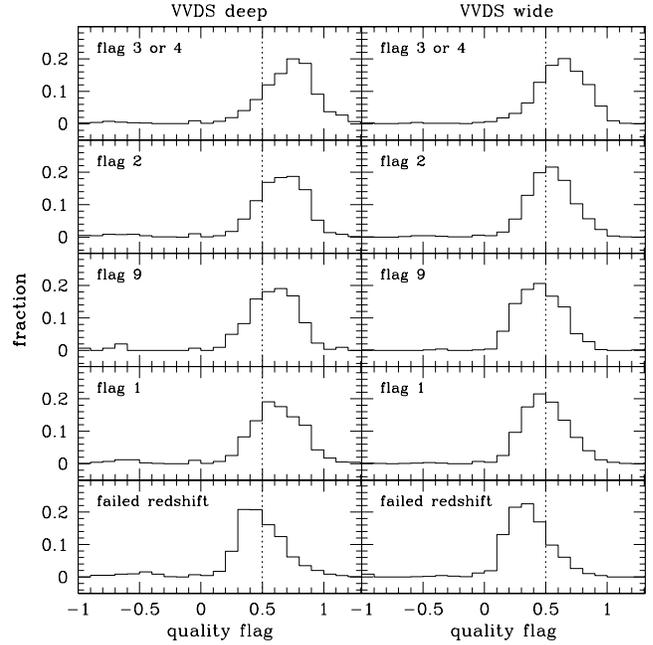}}
\caption{Distribution of the quality parameter for the VVDS Deep
  survey, divided by flag: the dotted line indicates the 0.5 value of
  the quality parameter
}
\label{qualityVsFlag_all}
\end{figure}
In Fig~\ref{qualityVsFlag_all}, for each flag, the distribution of the
quality parameter is shown for the VVDS Deep (left) and VVDS Wide 
(right) surveys. The dotted line indicates the 0.5 value of the 
quality parameter. As expected, more secure flags are assigned to 
objects showing, on average, a higher quality parameter, 
as shown by the peak of the histogram
moving towards higher values of quality 
going to more secure flags. An exception are the flag 9 objects, which 
show a distribution of the
quality parameter comparable to that of the flag 1
objects. This is not a surprise: we recall that a flag 9 is assigned when 
one secure single spectral feature in emission is visible, and, as
anticipated before, an emission line, if strong enough, can be
detected even in presence of low S/N continuum, or residual 
fringing patterns.\\
More quantitatively, in the VVDS Deep survey only
42\% of the failed spectra have a quality
parameter larger than 0.5, a percentage which goes down to 20\% for
the Wide survey. On the other hand, 82\% of the objects with a
very secure redshift flag ( flag 3 or 4) in the Deep survey 
are derived from spectra of
good quality (quality $>0.5$), a percentage which decreases to
73\% in the Wide case. \\
Thus, the quality parameter statistically strengthens the redshift
flag, and justifies it on the basis of the quality of the data. 
Furthermore, the coupling of the
two pieces of information allows one to easily select subsamples of objects for which not
only the redshift has the highest degree of reliability, but the data
are above a given quality and thus particularly suitable for detailed
studies of the continuum emission.
\section{General properties of the spectroscopic sample}
\begin{table*}
\caption{Statistics of redshift quality flags for VVDS Wide sample}
\label{redshiftTable}
\centering
\begin{tabular}{l  c    c     c     c     c     c    c    c   c    c    c    c    c   }
\hline\hline
                &\multicolumn{6}{l}{primary targets}&  \multicolumn{6}{l}{secondary targets}  &   \\   
Field/flag      & 0  & 1  & 2  & 3  & 4  & 9 & 20& 21& 22& 23&24&29&Total\\
\hline
VVDS-F02        &  38& 177& 506& 798&1501& 21& 49& 26& 44& 23&28& 6& 3217\\
  galaxies      & -  & 144& 434& 708&1121& 18&  -& 25& 41& 21&23& 4& 2539\\
  QSOs          & -  & 3  &   1&  12&  18&  3&  -&  0&  0&  0& 0& 2&   39\\
  stars         & -  & 30 &  71&  78& 362&  -&  -&  1&  3&  2& 5& -&  552\\
\\
VVDS-F10        & 327& 683& 916& 674& 961& 98& 85& 20& 21& 12&16& 8& 3821\\
  galaxies      & -  & 613& 772& 506& 413& 94&  -& 19& 18&  7& 8& 8& 2458\\
  QSOs          & -  & 6  & 15 & 17 & 3  &  4&  -&  0&  0&  0& 0& 0&   45\\
  stars         & -  & 64 & 129& 151& 545&  -&  -&  1& 3 &  5& 8& -&  906\\
\\
VVDS-F14        &240 &669 &1241&1268&1845&116&117& 59& 51& 19&32&12& 5669\\
  galaxies      & -  &572 & 935& 893& 915&105&  -& 55& 36& 16&13&12& 3552\\
  QSOs          & -  & 12 &   9&  14&   5& 11&  -&  0&  2&  0& 0& 0&   53\\
  stars         & -  & 85 & 297& 361& 925&  -&  -&  4& 13& 3 &19& 0& 1707\\
\\
VVDS-F22        &1507&2846&4783&3822&5671&499&377&153&122&106&95&46&20027\\
  galaxies      & -  &2504&3626&2357&1904&461& - &123& 99& 79&29&46&11228\\
  QSOs          & -  &22  &27  &36  &39  & 38& - & 1 & 1 & 1 & 2& 0&  167\\
  stars         & -  &320 &1130&1429&3728& - & - & 29& 22& 26&64& 0& 6748\\
\hline
\end{tabular}
\end{table*}
In Table ~\ref{redshiftTable} we summarize the statistics of redshift
measurement for the VVDS Wide sample. For reference, we also report
the redshift statistics for the VVDS Deep sample, once it is cut at a
limiting magnitude of $\mathrm{I_{AB}} \sim 22.5$. So far, in the 3
Wide survey fields we have accumulated 28166 spectra for primary
targets, including 16670 galaxies, 258 QSOs and 9164 stars. There are
only 2074 spectra for which the redshift measurement failed; this
corresponds to a success rate 
larger than 92\%.  
Secure redshift objects (flag between 2 and 9), are 21894,
almost 80\% of the sample. Although the magnitude limit is only
$\mathrm{I_{AB}}=22.5$, thanks to the large surveyed area ($\sim5.0\
deg^{2}$ of effective area), we have a fairly large sample of rare,
luminous galaxies at high
redshift: 979 with $1.0<z<1.4$ and 225 with $1.4<z<2.0$. The
highest secure redshift measured for a galaxy is 4.0573, while the
highest secure redshift object is a QSO at z=5.0163. On top of the
targeted sample, we also have 772 additional redshifts of objects
accidentally falling within the slit. Adding the data collected in
the F02 field limited to $\mathrm{I_{AB}}=22.5$, the VVDS Wide sample
comprises almost 20000 galaxies and 304 QSOs with measured redshift 
over 6.1 $\mathrm{deg}^{2}$.
\subsection{Magnitudes, sampling rate and redshift distribution}
\begin{figure}
\resizebox{\hsize}{!}{\includegraphics[clip=true]{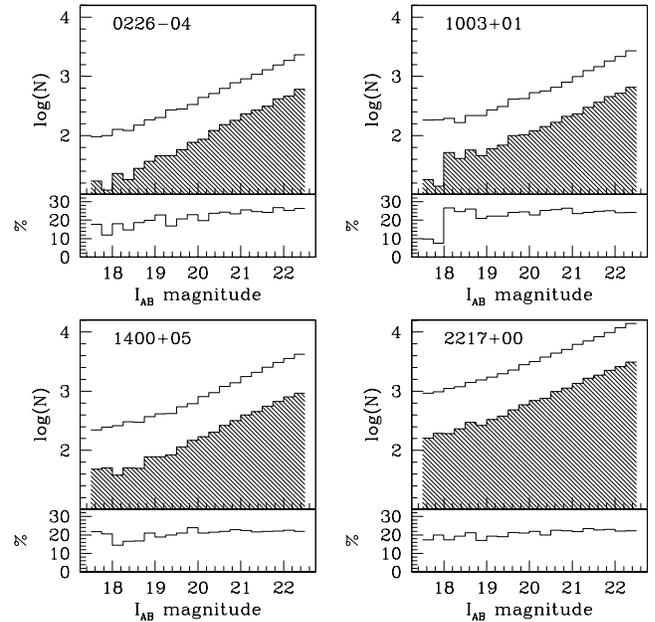}}
\caption{Spectroscopic survey sampling rate of the four fields as a
  function of magnitude: in the top panels, the apparent magnitude
  distribution for the parent photometric catalog (empty histogram)
  and the observed spectroscopic catalog (dashed histogram); in the bottom
  panels, the ratio of the two, corresponding to the effective sampling
  rate as a function of apparent magnitude.
}
\label{maghisto}
\end{figure}

\begin{figure}
\resizebox{\hsize}{!}{\includegraphics[clip=true]{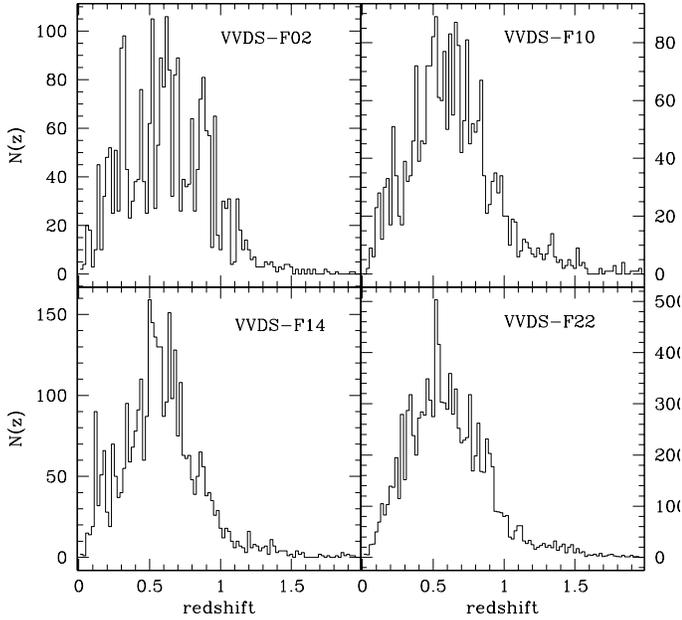}}
\caption{Observed Redshift distribution  to $\mathrm{I_{AB}}=22.5$
  in the four VVDS Wide fields, in redshift bins of $\Delta$$z=0.02$.
  All galaxies with quality flag between 1 to 19 have been used.}
\label{spechisto}
\end{figure}

In Fig.~\ref{maghisto}, the magnitude distribution of the photometric parent
catalog (top panels, empty histogram) and of the final
spectroscopic sample (top panels, shaded histogram)
is shown for the three VVDS Wide areas. For comparison, we show the
same plot for the VVDS-Deep F02 field, limited to $\mathrm{I_{AB}}=22.5$.
The bottom panels
show the fraction of observed over total objects vs. magnitude. In
all the three VVDS Wide areas, the fraction of observed objects is
between 20\%
and 25\%, over the full magnitude range,
very similar to the
sampling rate of the VVDS Deep area, once limited at $\mathrm{I_{AB}}=22.5$.
The slight trend
  favouring a better sampling at the faintest magnitude in the F02
  field is due to the intrinsic
  deeper limiting magnitude of the spectroscopic selection in this area,
  $I_{AB}<24$, which increases the probability of brighter objects to be
  discarded in favour of fainter ones \citep[see][]{vmmps}.\\
% maghisto
% macro maghisto.mac, comando mh 0
% usa file che comprendono TUTTI gli oggetti che hanno
% una misura di redhsift, qualunque sia l'area
In Fig.~\ref{spechisto}, the redshift distributions using all
available redshifts (irrespective of flags) for the four
different areas are shown. Table ~\ref{N_z_Table} shows that there
are no statistically significant differences between 
the N(z) obtained using all redshifts, and the
ones obtained using only secure redshifts (i.e. flag 2 to 9, 
redshift confidence
$>=80\%$).
% spechisto
% macro spechisto.mac, comando sh 0.02
%data summary table
\begin{table}
\tabcolsep0.3mm
\caption{Statistics on galaxy redshift distributions ($\mathrm{I_{AB}}<=22.5$)}
\label{N_z_Table}
\centering
\begin{tabular}{c c c c c c c}
\hline\hline
Field & 1$^{st}$ quartile & median &3$^{rd}$ quartile & 1$^{st}$ quartile
& median & 3$^{rd}$ quartile \\
      &\multicolumn{3}{c}{all flags}&\multicolumn{3}{c}{secure flags}     \\   
\hline
F02 & 0.374 & 0.611 & 0.834 & 0.369 & 0.611 & 0.824 \\
F10 & 0.386 & 0.616 & 0.856 & 0.368 & 0.605 & 0.855 \\
F14 & 0.351 & 0.560 & 0.724 & 0.333 & 0.559 & 0.767 \\
F22 & 0.404 & 0.571 & 0.810 & 0.382 & 0.560 & 0.770 \\
\hline
\end{tabular}
\end{table}
\subsection{Galaxy luminosities and stellar masses}
% magabsss and mass
% macro absMag.mac, comando Ab o AM
\begin{figure}
\resizebox{\hsize}{!}{\includegraphics[clip=true]{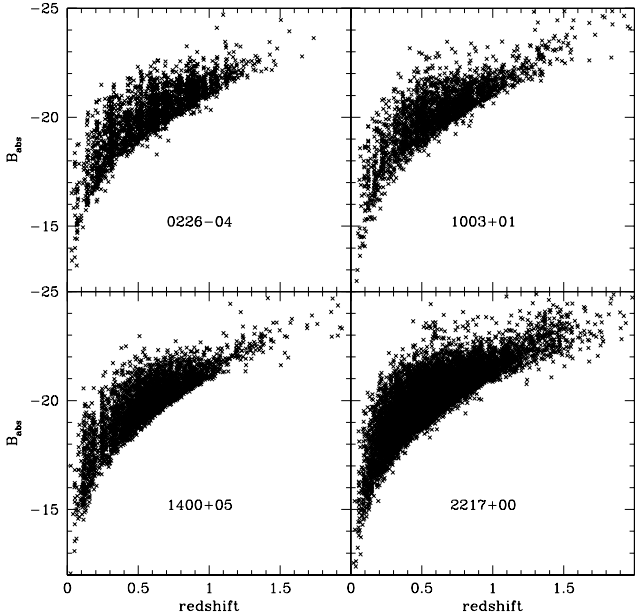}}
\caption{Absolute $B_{AB}$ magnitude vs. redshift in the four VVDS
  Wide areas.The F02 area has been cut to a limiting apparent magnitude $\mathrm{I_{AB}}=22.5$
}
\label{magabs}
\end{figure}
\begin{figure}
\resizebox{\hsize}{!}{\includegraphics[clip=true]{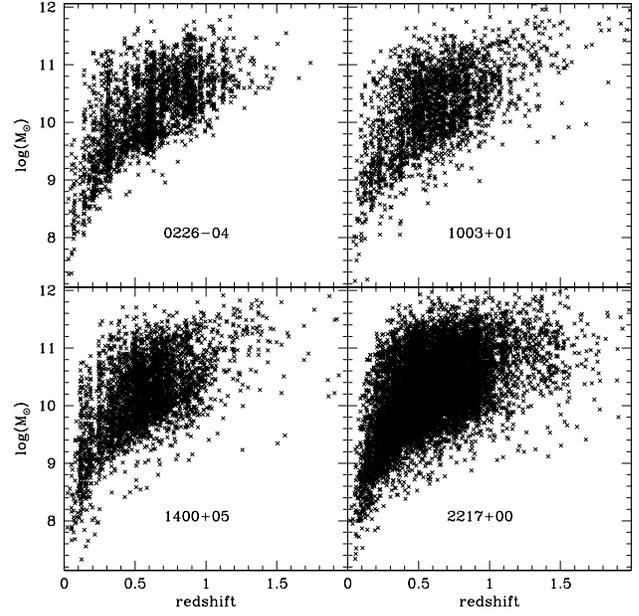}}
\caption{Stellar mass vs. redshift in the four VVDS Wide areas.The F02 area has been cut to a limiting apparent magnitude $\mathrm{I_{AB}}=22.5$.}
\label{mass}
\end{figure}
The large areas explored, coupled with the relative bright magnitude
limit,
make the VVDS Wide the ideal survey to explore the bright/massive ends
of the 
luminosity/mass function up to redshift $\sim 1$. As an example of the
potential of this sample for these studies, 
in Fig.~\ref{magabs} and ~\ref{mass}
we show the absolute B magnitude and stellar mass
vs. redshift distribution for the galaxies with secure redshifts in
the four areas. 
Absolute B magnitudes and stellar masses have been derived by fitting
the photometric and spectroscopic data with a grid of stellar
population synthesis models generated with the PEGASE2 population
synthesis code \citep{pegase}, and using the GOSSIP Spectral Energy 
Distribution
tool \citep{gossip}, where we have adopted a Salpeter IMF and
a delayed exponential SFH (see \citet{vvds_massfunc} for a thorough
discussion on the dependence of mass values on the different IMF adopted).
We can define a unique complete sample of
3542 bright galaxies with $M_{B_{AB}}<=-21$ up to $z\sim1$:
2136 galaxies in the F22 field, 412 in F10, 520 in F14, 474 in F02
(we remind that we have cut the F02
area to a limiting apparent magnitude $\mathrm{I_{AB}}=22.5$).
At the same limit in redshift, we have more than 11000 galaxies more
massive than log(\msun)=10.
(6547 galaxies in the F22 field, 1367 in F10, 2009 in F14, 1271 in F02), a sample which will allow a detailed study of
the properties of medium to high mass galaxies.
\subsection{A direct test of star-galaxy separation techniques}
As mentioned earlier, the VVDS was deliberately carried out
without any star-galaxy separation prior to spectroscopy.
When the survey was planned, only ground based $BVRI$ photometry was
available (and not over all fields), thus preventing us from using the
most efficient color based methods to discriminate between stars and
extragalactic objects.
Furthermore, the image quality of such ground based photometry was not
good enough to apply geometrical arguments discriminate between point-like and
extended sources down to the magnitude limits of the Deep and Wide
surveys.
Thus, we decided to follow the conservative approach of
not attempting any {\it a priori} removal of starlike objects based on
colors or compactness.
This has lead to the high stellar contamination of the spectroscopic
sample (up to $\sim$ 1/3 for the lower galactic latitude fields).
Using UKIDSS K photometry, and CFHTLS z photometry available in the
F22 and F02 field, we
can test with excellent statistics the 
performances of these 
star identification methods.
We 
thus applied to the spectroscopic sample the BzK criterium
described in \citet{Daddi}, coupled with a compactness criterium
based on the stellarity index provided by {\it Sextractor}: any
object with a stellarity index above or equal 0.9 is catalogued as
compact. An object is considered as a star if both
criteria are satisfied. 
To optimize the test, we used only objects
with secure redshift (redshift flag $>1$) and small photometric
errors (err$<0.1$) in the B, K and z Bands.\\
Applying this technique to the F02 data (which are
deeper and at high galactic latitude), we end up with a residual stellar
contamination of only $\sim 2\%$.  In the F22
field, which has a brighter magnitude limit and is located at lower
galactic latitude, the residual contamination
decreases from 35\% to 14\%. 
The price to be paid in terms
of galaxies which would have been {\it a priori} discarded is about
5\% in the F22 field, and about 2\% in the F02 case. 
We have checked
which kind of galaxies were typically discarded and found out that
they 
have the spectrum  of a normal elliptical galaxy.
Overall, we can state on the basis of observed data that the performance 
of these two coupled methods in discarding stars is highly efficient,
at the low price of a small loss of normal early type galaxies 
In addition, we note that we have applied the colour method
  in the standard
  form. Exploring in detail alternative color-color selections
  using the other available bands is beyond the scope of this
  paper. 
\section{Large-scale structure and density fluctuations in the VVDS
  Wide fields} 
\subsection{Galaxy spatial distribution in the F22 field}
\label{cone}
% cone diagram
% macro coneco_F22_slides.mac comando cone -0.5 1.0
% sotituire %%BoundingBox: 18 144 592 718 con
%%BoundingBox: 28 164 592 530
\begin{figure*}
\resizebox{\hsize}{!}{\includegraphics[clip=true]{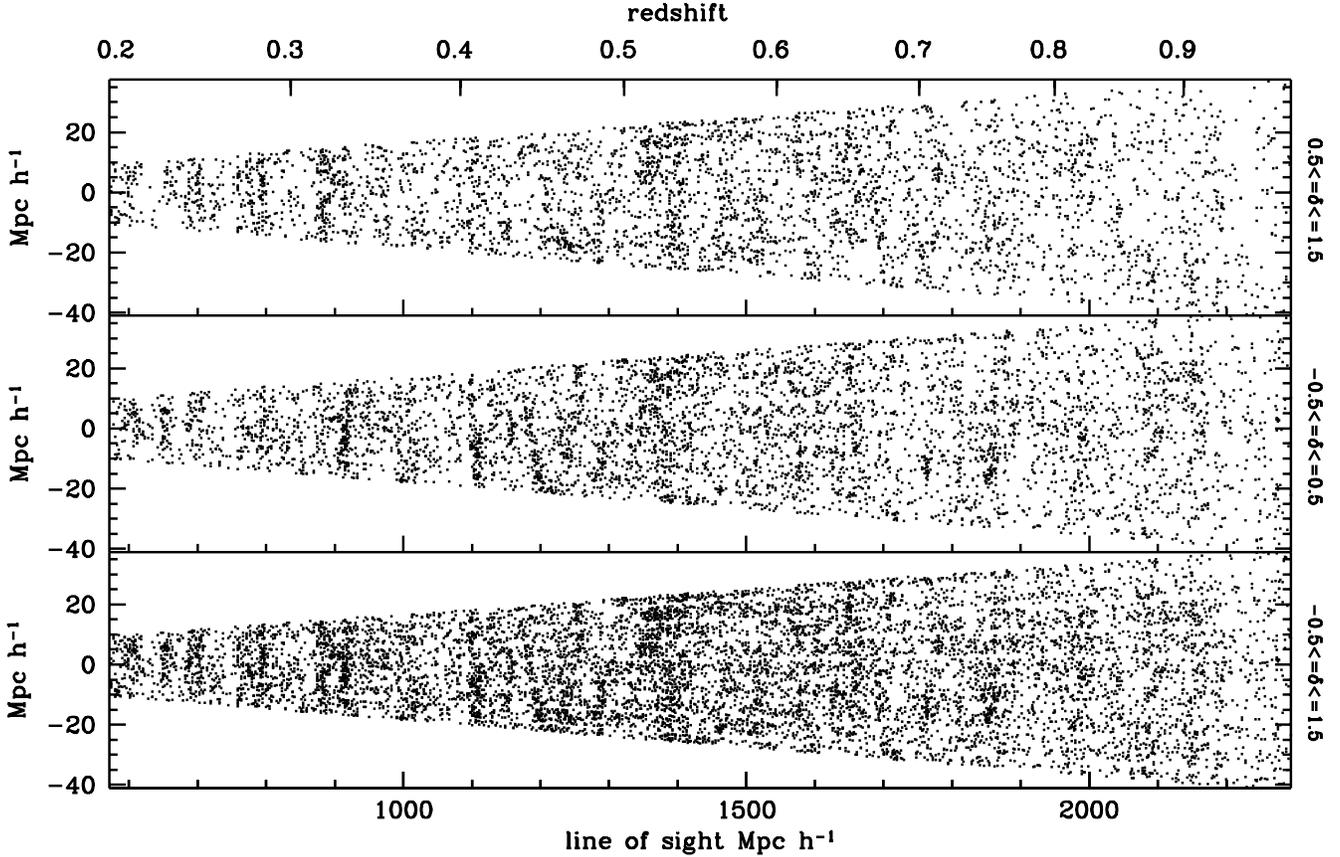}}
\caption{Cone diagrams of the 3D galaxy distribution in the 
the F22 field, projected on the right ascension plane for the whole
sample (lower panel) and for the two
1 deg slices in declination (upper panels)}
\label{cone_F22}
\end{figure*}
In Fig.~\ref{cone_F22} we show the redshift
space cone diagram of all
  galaxies observed in the F22 area, in co-moving coordinates and
  projected onto the right ascension plane.  The figure shows two
  declination slices, of 1 degree each, to better show the extension
  of the different structures.
Note that the aspect ratio is stretched along the vertical direction.
We can identify galaxy
overdensities at $z=0.28$, 0.33, 0.41, 0.53, 0.75, 0.82 and 0.9, some of
which extend over 
the full surveyed area, both in right ascension and declination:
at z=0.33 a very thin wall covers the
whole field of view of $24 \times 24~\mathrm{h^{-1}~Mpc}$; the structure at
$z\sim0.53$ is the most prominent and massive,
extending for almost 80 $\mathrm{h^{-1}~Mpc}$ along the line
of sight, and 40 $\mathrm{h^{-1}~Mpc}$ across. Its presence
strongly influences the redshift distribution in this field,
lowering its median value and steepening its rise at low redshifts.
Such ``thick wall'' has several subconcentrations, better visible in the
slices in declination of Fig.~\ref{cone_F22}.
The other visible structures look rather more compact, with a comoving transverse size of the order of 20
$\mathrm{h^{-1}~Mpc}$, and confined within the 2
square degrees.
\subsection{Mean redshift distribution up to $\mathrm{I_{AB}}=22.5$}
% macro fig8b.mac, comando univ
\begin{figure}
\resizebox{\hsize}{!}{\includegraphics[clip=true]{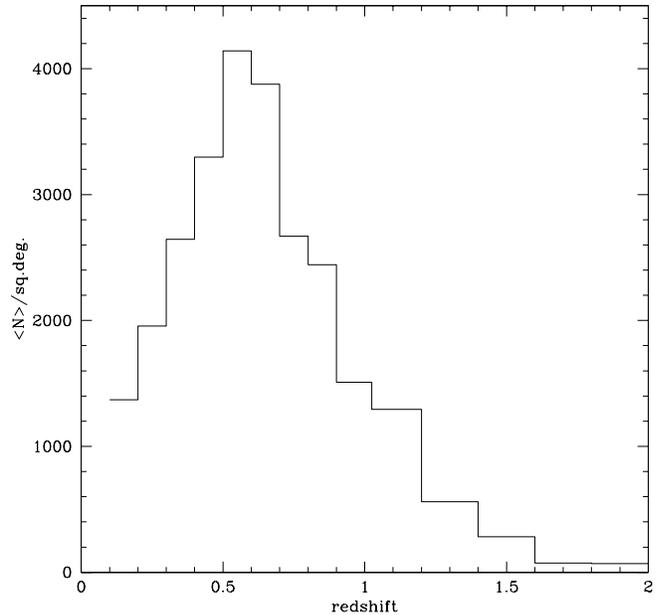}}
\caption{Mean redshift distribution per square degree obtained in
  the full survey area of 6.1 square degrees}
\label{univ}
\end{figure}
In terms of their broad shape and peak position, the galaxy
redshift distributions in the four areas
are relatively similar.
At the same time, however, significant field-to-field
variations are evident (e.g. the thick wall at 0.53 in the F22 field,
as outlined in the previous section).  In this and the following
sections we 
quantify
this variance and compare it to theoretical expectations, as obtained both from
the observed two-point correlation function and from mock surveys
built using numerical/semi-analytic models.\\
Combining the four fields, appropriately taking into account
the effective area and the sampling rate of each field, we
can derive our current best estimate of the redshift distribution 
of a magnitude selected sample to 
$\mathrm{I_{AB}}<=22.5$.
The result is shown in Figure~\ref{univ}
and the corresponding values are reported in Table \ref{univ_table}
for convenience. In this figure and table, we
use a binning of $\Delta$z=0.1 up to z=1,
and 0.2 at higher redshift,
in order
to smooth out the smaller structures present in the different fields.
This represents the most
accurate redshift distribution mesured to date at these faint
magnitudes,  
based on $\sim 20,000$ galaxies over a total area of
6.1 deg$^2$, and it can provide an important reference for galaxy formation
models. 

% cosmic variance
% macro cosm_var, comando all2  5 14
% last point is 0.2 in redshift
% quindi macro fig8b.mac, comando 8b
\begin{figure}
\resizebox{\hsize}{!}{\includegraphics[clip=true]{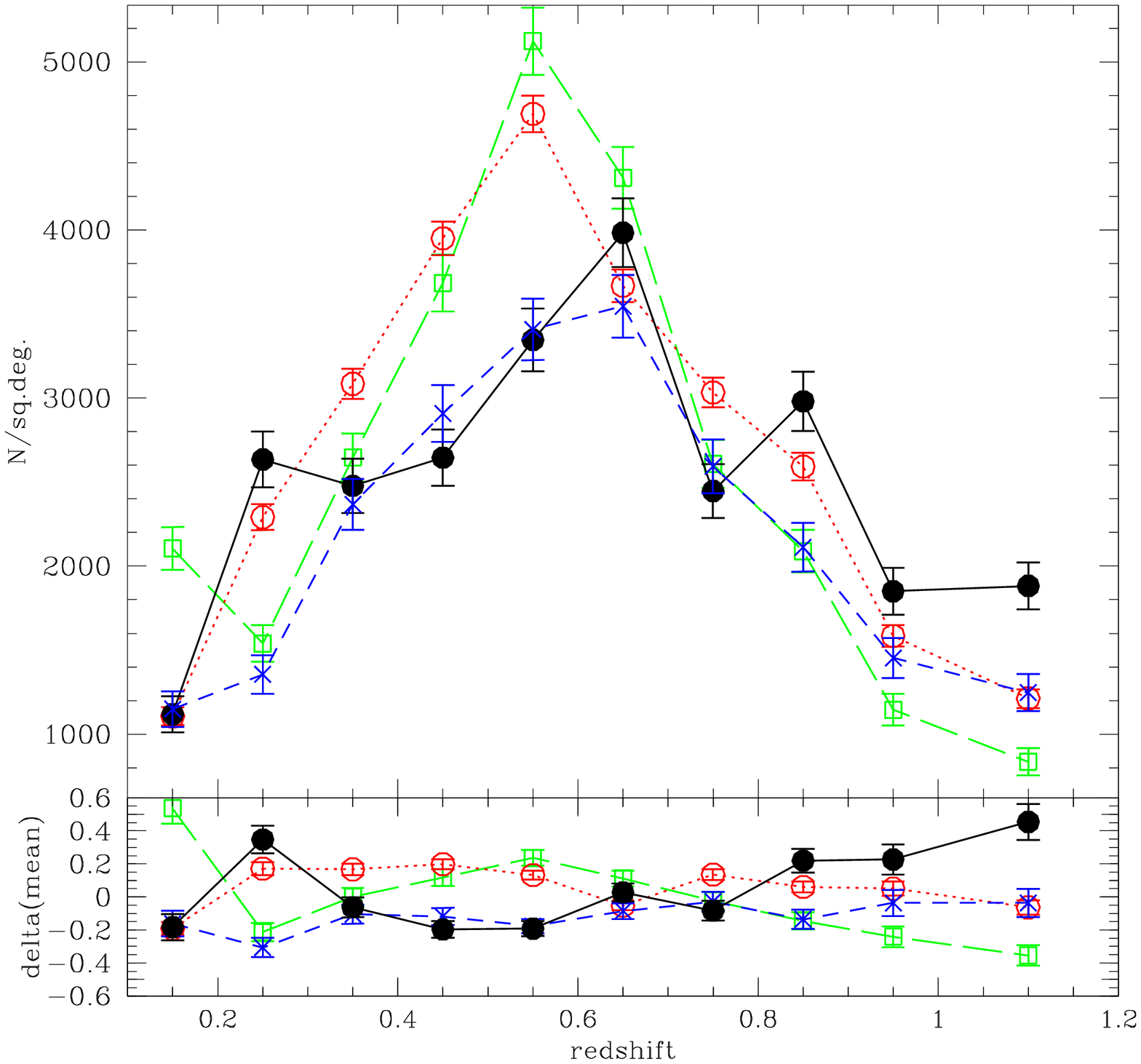}}
\caption{Top: Redshift distribution (in galaxies per unit effective
  area) in the four VVDS-Wide areas:    F02, black dots and solid
  line, F10 blue crosses and short dashed   line, F14 green squares
  and long  dashed line, and F22, red circles and dotted line.  The
  sampling corrections have been assumed to be independent of
  redshift. Bottom: field-to-field variations, relative to the
  globally averaged redshift distribution of figure~\ref{univ}.}
\label{cosm_var}
\end{figure}
% build table running macro fig8b
\begin{table}
\tabcolsep0.3mm
\caption{Galaxy surface density as a function of redshift up to a limiting mag $\mathrm{I_{AB}}<=22.5$
averaged over 6.1 square degrees}
\label{univ_table}
\centering
\begin{tabular}{l c c c }
\hline\hline
z ~~~~& mean~~ & max~~~ & min ~~~ \\
      &~~$gal/deg^2$~~&~$gal/deg^2$~~&~$gal/deg^2$~~ \\
\hline
0.15  & 1370   &  2105 & 1107  \\
0.25  & 1956   &  2635 &  1356  \\ 
0.35  & 2644   &  3085 &   2367 \\  
0.45  & 3296   &  3950 &  2645 \\
0.55  & 4142   &  5123 &  3346 \\
0.65  & 3877   &  4310 &  3546 \\  
0.75  & 2670   &  3033 &  2447 \\
0.85  & 2443   &  2980 &  2089 \\
0.95  & 1509   &  1851 &  1145 \\  
1.1   & 1294   &  1882 &   836 \\ 
1.3   & 561   &   746 &   439 \\
1.5   &  281   &   357 &   193 \\   
1.7   &  74   &   113 &    31 \\   
1.9   &  68   &   108 &    31 \\
\hline
\hline
\end{tabular}
\end{table}
\subsection{Field to field variations}
With this unprecedented area surveyed, it becomes
possible to quantify the variations in each of the four
fields with respect to this average distribution. This is shown
in Fig.~\ref{cosm_var}. The top panel reports the redshift
distribution of the four fields, using a $\Delta$z=0.1 binning.
For reference, around the peak of the distribution z=[0.5,0.6]
such a redshift bin corresponds to a comoving radial size of 222
$\mathrm{h^{-1}~Mpc}$.  Error bars correspond to Poissonian
errors. In the bottom panel of figure \ref{cosm_var}, we show
the fractional difference between the observed N(z) in each field, and the
average distribution. This comparison of the fluctuations in the
different fields for fixed redshift bins is inevitably
qualitative.  In fact, given the very different areas covered
in the four fields, the same redshift range corresponds to rather
different volumes.
For example, the smallest field, F02, covers 0.5 deg$^2$, i.e. 8
times smaller than the largest one, F22 (4 deg$^2$). This is
certainly one reason for the higher variance in the F02 field
(Fig.~\ref{cosm_var}, filled circles).\\
% surface density
% macro plotta_conta.mac comando pc 
% Se anche mock qiondi pc_mock.mac comando pcm quindi erase, pcm, pc
% dev postencap surface_density.ps 
\begin{figure}
\centering
{\includegraphics [bb=26 150 306 700, width=9cm,clip] {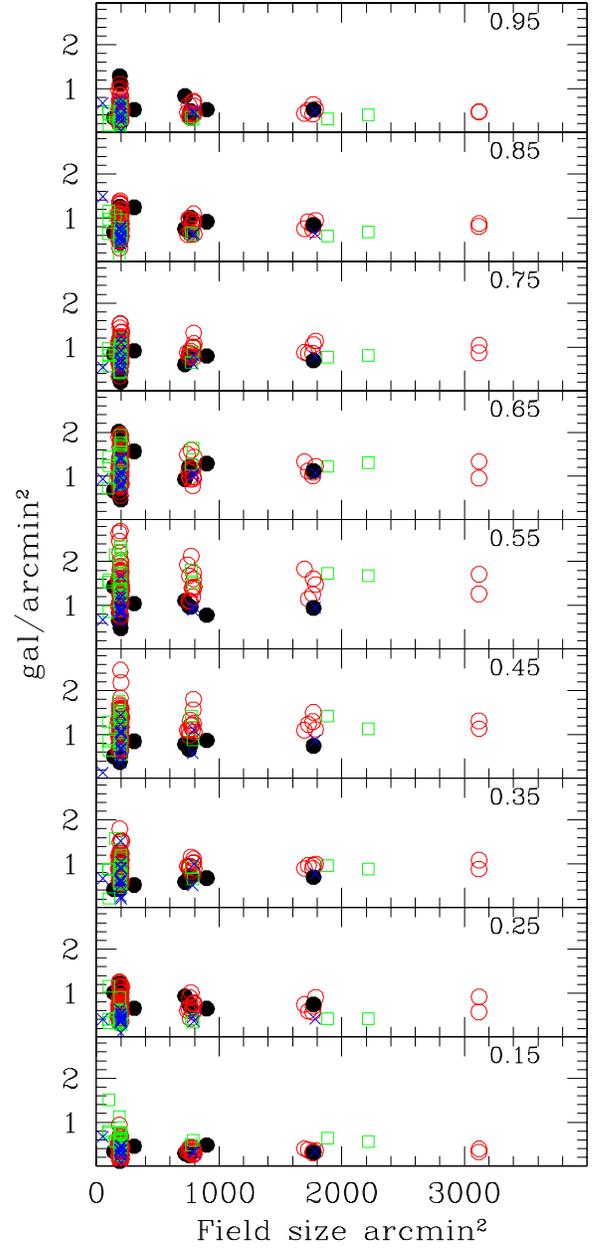}}
\caption{A different view of the galaxy redshift distribution,
  evidencing the effect of varying the field size.  The ascissa
  shows the size of the field on the sky in square arcminutes, while
  the ordinate gives the corresponding value of the surface density
  (in galaxies per unit area) in the given redshift bin of size
  $\Delta$$z=0.1$ (centered at the mean
  redshift given in each panel).  Sub-fields of increasing size are
  coded according to their parent VVDS-Wide field: F02
  (black dots), F10 (blue crosses) F14 (green 
  squares) and F22 (red circles).  Clearly, the largest-size sub-fields
  can only be drawn from the largest parent field, i.e. F22.  
Statistical errors on
  each measurement are of the same size as the symbols.
 }
\label{surface_density}
\end{figure}
{To properly estimate the intrinsic variance as a function of
scale, we have therefore defined a set of square sub-fields over the
four survey areas, with increasing angular size. The variance is
then computed among the set of N$_i$ homogeneous volumes having
identical size on the sky and along the redshift direction. The
result is summarized in Fig.~\ref{surface_density}. In practice,
each sky region (represented by a different color and symbol) has
been divided -- for a given size -- into the largest possible number
of subareas that could be accomodated.
Galaxy densities have been computed in each sub-field and for
different redshift bins, properly correcting for the average
sampling of the area. Table \ref{sv_table} shows quantitavely the
results illustrated in Fig.~\ref{surface_density}. For each area
size, and each redshift bin, we give the number of sub-fields of that size available,
the median value,
upper/lower quartiles and maximum and minimum of the observed galaxy
density.
For the largest area, where only 2 measurements are available, we
computed the arithmetic mean instead of the median. For the smaller
scales represented (190 arcmin$^{2}$) the observed large
fluctuations (up to a factor of four between adjacent areas) are not
surprising, as we are essentially looking at scales of the order of
a few Mpc. At redshifts approaching unity, and for large angular
sizes (30 arcmin correspond to 10 $\mathrm{h^{-1}~Mpc}$ at this
redshift) the spread can still be a factor of 2, an indication that
important structures exist and are not uncommon at such redshift.
It is interesting to see, using the larger areas, how much variance
we expect in a field of $0.5\deg^{2}$ (1800 square arcminutes) like
F02, i.e. the field of the VVDS-Deep survey. For example, at
redshift 0.75 we still see peak to peak fluctuations of $\sim$ 30 \% in
the counts. We also notice a significant excess fluctuation in the data
from the field F22 (red circles) in the redshift bin 0.5-0.6. In
particular, all counts are shifted towards higher values, reflecting
the presence of the global large-scale fluctuation covering the full
field already mentioned in section \ref{cone}.}\\
%% NEW TABLE
\begin{table}
\tabcolsep0.99mm
\caption{Galaxy density median values and spread in different area sizes at different
  redshifts}
% ptinted out by plotta_conta.mac
\label{sv_table}
\centering
\begin{tabular}{c c c c c c c c}
\hline\hline
$<z>$& area & N areas & median & upper & lower & max & min \\
    & arcmin$^2$ & & & quartile &quartile  & & \\
\hline
0.15 & 190 &96 & 0.37 & 0.52 & 0.25 & 0.67 & 0.18 \\
0.25 & 190 &96 & 0.61 & 0.82 & 0.45 & 1.07 & 0.35 \\
0.35 & 190 &96 & 0.86 & 1.05 & 0.62 & 1.23 & 0.50 \\
0.45 & 190 &96 & 1.03 & 1.38 & 0.77 & 1.57 & 0.60 \\
0.55 & 190 &96 & 1.32 & 1.63 & 1.01 & 1.98 & 0.79 \\
0.65 & 190 &96 & 1.10 & 1.43 & 0.94 & 1.70 & 0.79 \\
0.75 & 190 &96 & 0.82 & 1.05 & 0.67 & 1.19 & 0.50 \\
0.85 & 190 &96 & 0.74 & 0.96 & 0.59 & 1.15 & 0.51 \\
0.95 & 190 &96 & 0.45 & 0.58 & 0.32 & 0.75 & 0.25 \\
\hline
0.15 & 780 &19 & 0.34 & 0.42 & 0.29 & 0.48 & 0.26 \\
0.25 & 780 &19 & 0.66 & 0.76 & 0.45 & 0.93 & 0.36 \\  
0.35 & 780 &19 & 0.92 & 1.00 & 0.69 & 1.12 & 0.59 \\  
0.45 & 780 &19 & 1.10 & 1.26 & 0.87 & 1.57 & 0.66 \\  
0.55 & 780 &19 & 1.38 & 1.67 & 0.98 & 1.92 & 0.89 \\
0.65 & 780 &19 & 1.14 & 1.29 & 0.98 & 1.59 & 0.93 \\
0.75 & 780 &19 & 0.81 & 0.91 & 0.76 & 1.08 & 0.62 \\
0.85 & 780 &19 & 0.75 & 0.94 & 0.63 & 1.01 & 0.62 \\
0.95 & 780 &19 & 0.45 & 0.61 & 0.38 & 0.72 & 0.35 \\
\hline
0.15 & 1725 &8 & 0.36 & 0.40 & 0.33 & 0.64 & 0.30\\
0.25 & 1725 &8 & 0.70 & 0.74 & 0.56 & 0.91 & 0.41\\  
0.35 & 1725 &8 & 0.96 & 0.97 & 0.89 & 1.00 & 0.70\\  
0.45 & 1725 &8 & 1.24 & 1.42 & 1.10 & 1.51 & 0.75\\  
0.55 & 1725 &8 & 1.45 & 1.73 & 1.15 & 1.83 & 0.94\\
0.65 & 1725 &8 & 1.12 & 1.23 & 1.09 & 1.33 & 1.01\\
0.75 & 1725 &8 & 0.85 & 1.05 & 0.77 & 1.14 & 0.69\\
0.85 & 1725 &8 & 0.83 & 0.91 & 0.74 & 0.94 & 0.59\\
0.95 & 1725 &8 & 0.50 & 0.53 & 0.44 & 0.64 & 0.31\\
\hline
0.15 & 3120 &2 & 0.35 & - & - & 0.39 & 0.32\\
0.25 & 3120 &2 & 0.74 & - & - & 0.91 & 0.57\\  
0.35 & 3120 &2 & 0.98 & - & - & 1.08 & 0.88\\  
0.45 & 3120 &2 & 1.22 & - & - & 1.31 & 1.13\\  
0.55 & 3120 &2 & 1.48 & - & - & 1.71 & 1.26\\
0.65 & 3120 &2 & 1.15 & - & - & 1.34 & 0.96\\
0.75 & 3120 &2 & 0.95 & - & - & 1.04 & 0.87\\
0.85 & 3120 &2 & 0.83 & - & - & 0.87 & 0.80\\
0.95 & 3120 &2 & 0.47 & - & - & 0.47 & 0.47\\
\hline
\end{tabular}
\end{table}
\subsection{Measuring cosmic variance}
%% NEW SECTION
% surface density
% macro calcola.mac comando c 1 3
% dev postencap cosmic_variance.ps 
\begin{figure}
\centering
{\includegraphics [bb=25 150 306 700, width=9cm,clip]{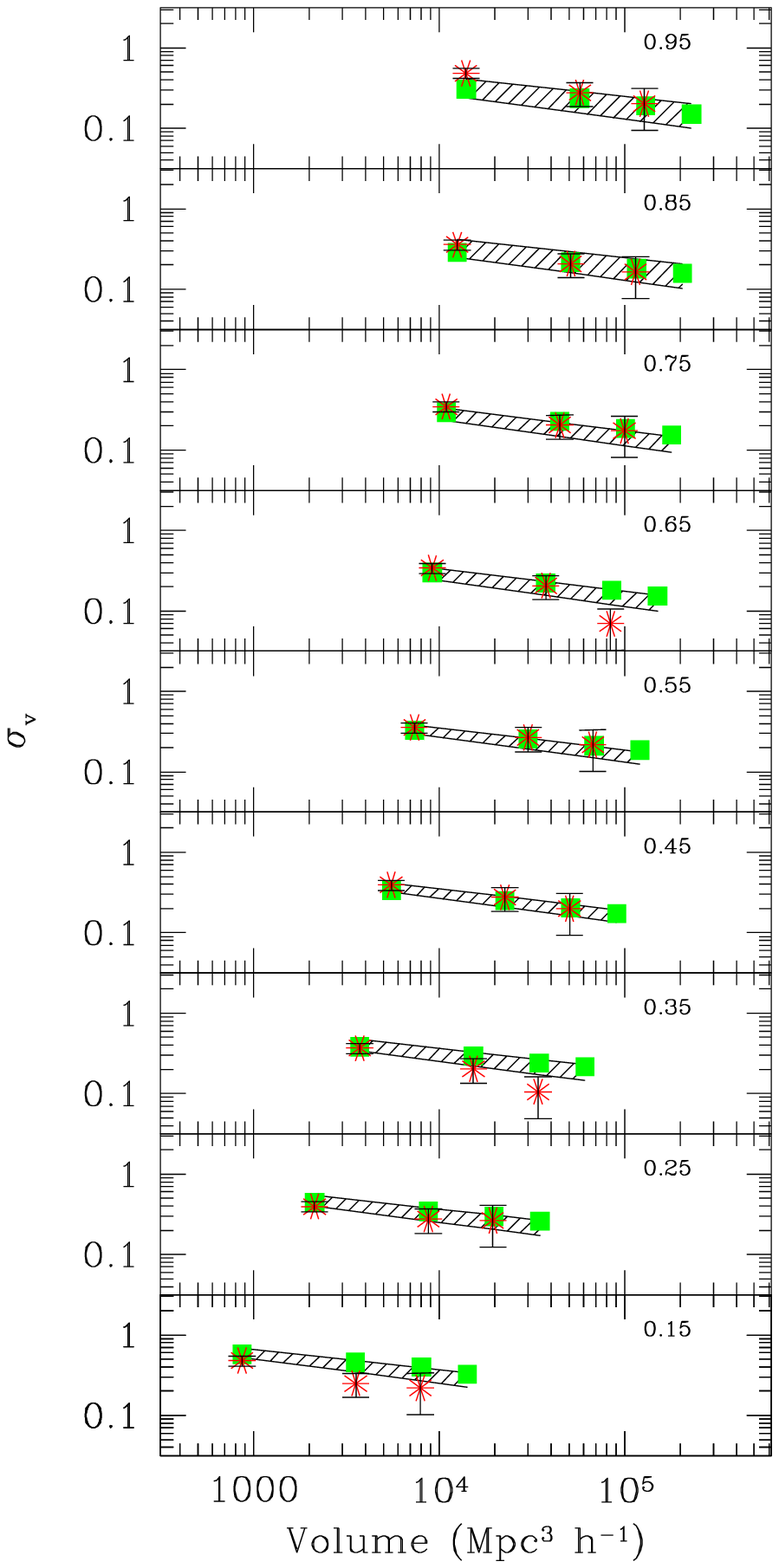}}
\caption{Comparison of the {\it rms} number density fluctuation (the
square-root of the variance) among independent sub-areas of different
size drawn from the four VVDS-Wide survey fields (red asterisks), with
that predicted on the basis of the galaxy two-point correlation
function (dashed bands). This is shown as a function of redshift (see
insets). The green filled squares give the same
directly measured {\it rms} fluctuations from the Millennium
mock samples.
}
\label{cosmic_variance}
\end{figure}
The results of Fig.~\ref{surface_density} can be translated from the
observational space into a framework which is theoretically easier to
interpret in terms of cosmic variance and expectations from galaxy
clustering. Given a set of N identical volumes with volume $V$, we can
define the observed variance among them as
\\
\begin{equation}
\sigma_{v}^{2} = \frac{<N^{2}>-<N>^{2}}{<N>^{2}}-\frac{1}{<N>}\\
\label{equation1}
\end{equation}
(e.g. \citet{somerville}), where the last term is the correction for
Poissonian shot noise. In the following we will compute $\sigma\mathrm{_{v}^{2}}$
following eq. \ref{equation1} only when the Poisson shot noise is
smaller than 10\%.
The observed variance in the counts at a given redshift 
can be compared to that expected from
the two-point correlation function of the galaxy sample. Following \citep{peebles}
\begin{equation}
\sigma_{v}^{2} = {1\over V^2} \int_Vd^3x_1d^3x_2
\xi\left(|\vec{x_1}-\vec{x_2}|\right)
\end{equation}
If the galaxy correlation function can be described as a power law,
$\xi(r)=(r/r_0)^{-\gamma}$,  then this expression
becomes 
\begin{equation}
\sigma_{v}^{2} = J_{2}(r_{0}/r)^{\gamma}
\label{equation3}
\end{equation}
where 
$J_{2}=72.0/[(3-\gamma)(4-\gamma)(6-\gamma)2^{\gamma}]$ and $\mathrm
r_{0}$ and $\gamma$ are measured from the observations.  

Following the same approach as in \citet{vvds_clus},
and using the VVDS-Deep F02 data limited at
$\mathrm{I}_{\mathrm{AB}}\le 22.5$, we have estimated
the best fit correlation function parameters in different redshift
bins, and used eq. \ref{equation3} to check whether the observed 
variance measured from the field-to-field scatter (as from eq.
\ref{equation1}) can be recovered
consistently by extrapolating the correlation function measured from a
much smaller field.  The results are shown in Fig.
\ref{cosmic_variance}, where we plot the observed square-root
$\sigma_V$ of the variance (i.e. the value of the {\it rms}
fluctuation) against the volume. The red asterisks correspond to the
direct measurement, obtained from the scatter among N$_i$ samples within
the given volume. The dashed area shows the same quantity as obtained
using the 3$\sigma$ confidence intervals of the VVDS-Deep F02
correlation function (limited at
$\mathrm{I}_{\mathrm{AB}}\le 22.5$).\\
One notices immediately that the variance directly estimated from the
galaxy counts in the different fields is 
in excellent agreement with the cosmic variance
as estimated from the correlation function.
Only at redshift 0.35 
and 0.65 and for volumes of $\sim 10^5$
h$^{-1}~Mpc^{3}$ , field to field
variance appears smaller than the one predicted from the correlation
function parameters.
Looking back at the distribution of the
number counts (fig. \ref{cosm_var}), at these two redshifts we note
a remarkable similarity among the different fields, as well as 
in the galaxy surface density distribution in
fig. \ref{surface_density} and table \ref{sv_table}.
This similarity automatically converts in a lower field to field
variance, which in any case remains compatible with the one computed
from correlation function at a $1.2\sigma$ level.

Using 100 quasi-independent mock samples of $2\times 2$ degrees
\citep[for details, see][]{guzzo_nature} built applying to the Millennium
simulation \citep{millennium,momaf} the semi-analytic prescription of
\citet{delucia}, we have computed
model predictions for cosmic variance again using eq.~\ref{equation1}. The 
model predictions (green filled squares) are quite well consistent
with the observed field-to-field variance, with a difference which is
at most $1.5\sigma$ at z=0.65 for volumes of $\sim 10^5$
h$^{-1}~Mpc^{3}$ .

\section{Public Data Release and Database Access}
We are publicly releasing all redshift measurements in the F22 area
through the CENCOS 
(CENtre de COSmologie) database environment on our web site
http://cencosw.oamp.fr with access to the database built under the
Oracle environment, and through VO services ({\it VVDS\_WIDE} ConeSearch
service).  The catalog can be searched by coordinates, redshift
interval, identification number, in combination with the spectra
quality flags. 
Spectra in FITS format are already available on the same site (or
through VO SSA service) for the F02 area, both from the CENCOS site
and
the VO SSA service {\it VVDS\_F02\_DEEP}.
The remaining redshifts, together with all spectra in FITS format,
will be available as soon as the whole set of available data will be
measured.
\section{Summary}
The VVDS Wide survey is still ongoing but it has already measured
redshifts for 26864 objects (including serendipitous objects) 
in 3 areas covering a total of 5.6
$\deg^{2}$ to a limiting magnitude of $\mathrm{I_{AB}}=22.5$, to which we 
can add 3130 redshifts to the same limiting magnitude obtained by
the VVDS Deep survey in the F02 field. The
success rate in redshift measurement is more than 92\% and more than three
quarters of the redshifts
have a confidence higher than 80\%. Overall (i.e. including the F02 field
and serendipitous objects) the current sample includes 19777 galaxies,
304 broad
line QSOs, and 9913 stars, while the total area covered amounts to
6.1 $\deg^{2}$. When completed, the total area coverage will be of 8.6
$\deg^{2}$, and the total number of redshifts of the order of 50000.
\\
The large number of
redshifts available in the F22 field, coupled with a sampling rate of 
 $\sim$23\%, allows to identify and describe several prominent
structures present along the line of sight up to 2500 $\mathrm{h^{-1}~Mpc}$.
Typical sizes are of the order of 20 $\mathrm{h^{-1}~Mpc}$, but one large clumpy
structure extends for almost 80 $\mathrm{h^{-1}~Mpc}$ along the line
of sight, and 40 $\mathrm{h^{-1}~Mpc}$ across. \\
We give the mean N(z) distribution averaged over 6.1
$\deg^{2}$ (fig. \ref{univ}) for a sample limited in magnitude to
$\mathrm{I_{AB}}=22.5$. We have estimated field to field variations in
terms of number counts (see fig. \ref{cosm_var} and table \ref{univ_table}) and
galaxy surface density both as a function
of field size and redshift (see fig. \ref{surface_density} and table 
\ref{sv_table}), showing
that differences as high as a factor of two can 
exist at z=1 for still relatively large scales of the
order 30 arcminutes, like those considered in many deep surveys today. 
For fields limited to smaller scales (of the order
10 arcminutes), the spread in galaxy densities can be up to a factor 2.5.
Still, the observed cosmic
variance is consistent both with what derived from the correlation function
parameters, and from theoretical simulations (see fig.\ref{cosmic_variance}).\\
In  
addition to the evolution of clustering and large-scale structure,  
this data set is best suited to study in detail the bright end of the
luminosity function, as well as the massive end of the mass function, 
up to $z\sim1$ in four different fields observed with identical
purely magnitude limited selection. \\
The redshift catalog for the F22 area is available at the web site
http://cencosw.oamp.fr,
or via ConeSearch VO service {\it VVDS\_WIDE}.

\begin{acknowledgements}
This research has been developed within the framework of the VVDS
consortium.\\
This work has been partially supported by the
CNRS-INSU and its Programme National de Cosmologie (France),
and by Italian Ministry (MIUR) grants
COFIN2000 (MM02037133) and COFIN2003 (num.2003020150) 
and by INAF grants (PRIN-INAF 2005).\\
The VLT-VIMOS observations have been carried out on guaranteed
time (GTO) allocated by the European Southern Observatory (ESO)
to the VIRMOS consortium, under a contractual agreement between the
Centre National de la Recherche Scientifique of France, heading
a consortium of French and Italian institutes, and ESO,
to design, manufacture and test the VIMOS instrument.
\end{acknowledgements}

\bibliographystyle{aa}
\bibliography{paper}

\begin{thebibliography}{49}
\expandafter\ifx\csname natexlab\endcsname\relax\def\natexlab#1{#1}\fi

\bibitem[{{Abazajian} {et~al.}(2003){Abazajian}, {Adelman-McCarthy},
  {Ag{\"u}eros}, {Allam}, {Anderson}, {Annis}, {Bahcall}, {Baldry}, {Bastian},
  {Berlind}, {Bernardi}, {Blanton}, {Blythe}, {Bochanski}, {Boroski},
  {Brewington}, {Briggs}, {Brinkmann}, {Brunner}, {Budav{\'a}ri}, {Carey},
  {Carr}, {Castander}, {Chiu}, {Collinge}, {Connolly}, {Covey}, {Csabai},
  {Dalcanton}, {Dodelson}, {Doi}, {Dong}, {Eisenstein}, {Evans}, {Fan},
  {Feldman}, {Finkbeiner}, {Friedman}, {Frieman}, {Fukugita}, {Gal},
  {Gillespie}, {Glazebrook}, {Gonzalez}, {Gray}, {Grebel}, {Grodnicki}, {Gunn},
  {Gurbani}, {Hall}, {Hao}, {Harbeck}, {Harris}, {Harris}, {Harvanek},
  {Hawley}, {Heckman}, {Helmboldt}, {Hendry}, {Hennessy}, {Hindsley}, {Hogg},
  {Holmgren}, {Holtzman}, {Homer}, {Hui}, {Ichikawa}, {Ichikawa}, {Inkmann},
  {Ivezi{\'c}}, {Jester}, {Johnston}, {Jordan}, {Jordan}, {Jorgensen},
  {Juri{\'c}}, {Kauffmann}, {Kent}, {Kleinman}, {Knapp}, {Kniazev}, {Kron},
  {Krzesi{\'n}ski}, {Kunszt}, {Kuropatkin}, {Lamb}, {Lampeitl}, {Laubscher},
  {Lee}, {Leger}, {Li}, {Lidz}, {Lin}, {Loh}, {Long}, {Loveday}, {Lupton},
  {Malik}, {Margon}, {McGehee}, {McKay}, {Meiksin}, {Miknaitis}, {Moorthy},
  {Munn}, {Murphy}, {Nakajima}, {Narayanan}, {Nash}, {Neilsen}, {Newberg},
  {Newman}, {Nichol}, {Nicinski}, {Nieto-Santisteban}, {Nitta}, {Odenkirchen},
  {Okamura}, {Ostriker}, {Owen}, {Padmanabhan}, {Peoples}, {Pier}, {Pindor},
  {Pope}, {Quinn}, {Rafikov}, {Raymond}, {Richards}, {Richmond}, {Rix},
  {Rockosi}, {Schaye}, {Schlegel}, {Schneider}, {Schroeder}, {Scranton},
  {Sekiguchi}, {Seljak}, {Sergey}, {Sesar}, {Sheldon}, {Shimasaku}, {Siegmund},
  {Silvestri}, {Sinisgalli}, {Sirko}, {Smith}, {Smol{\v c}i{\'c}}, {Snedden},
  {Stebbins}, {Steinhardt}, {Stinson}, {Stoughton}, {Strateva}, {Strauss},
  {SubbaRao}, {Szalay}, {Szapudi}, {Szkody}, {Tasca}, {Tegmark}, {Thakar},
  {Tremonti}, {Tucker}, {Uomoto}, {Vanden Berk}, {Vandenberg}, {Vogeley},
  {Voges}, {Vogt}, {Walkowicz}, {Weinberg}, {West}, {White}, {Wilhite},
  {Willman}, {Xu}, {Yanny}, {Yarger}, {Yasuda}, {Yip}, {Yocum}, {York},
  {Zakamska}, {Zehavi}, {Zheng}, {Zibetti}, \& {Zucker}}]{sdss}
{Abazajian}, K., {Adelman-McCarthy}, J.~K., {Ag{\"u}eros}, M.~A., {et~al.}
  2003, \aj, 126, 2081

\bibitem[{{Arnouts} {et~al.}(2005){Arnouts}, {Schiminovich}, {Ilbert},
  {Tresse}, {Milliard}, {Treyer}, {Bardelli}, {Budavari}, {Wyder}, {Zucca}, {Le
  F{\` e}vre}, {Martin}, {Vettolani}, {Adami}, {Arnaboldi}, {Barlow},
  {Bianchi}, {Bolzonella}, {Bottini}, {Byun}, {Cappi}, {Charlot}, {Contini},
  {Donas}, {Forster}, {Foucaud}, {Franzetti}, {Friedman}, {Garilli},
  {Gavignaud}, {Guzzo}, {Heckman}, {Hoopes}, {Iovino}, {Jelinsky}, {Le Brun},
  {Lee}, {Maccagni}, {Madore}, {Malina}, {Marano}, {Marinoni}, {McCracken},
  {Mazure}, {Meneux}, {Merighi}, {Morrissey}, {Neff}, {Paltani}, {Pell{\` o}},
  {Picat}, {Pollo}, {Pozzetti}, {Radovich}, {Rich}, {Scaramella}, {Scodeggio},
  {Seibert}, {Siegmund}, {Small}, {Szalay}, {Welsh}, {Xu}, {Zamorani}, \&
  {Zanichelli}}]{vvds_imaging_uv}
{Arnouts}, S., {Schiminovich}, D., {Ilbert}, O., {et~al.} 2005, \apjl, 619, L43

\bibitem[{{Blaizot} {et~al.}(2005){Blaizot}, {Wadadekar}, {Guiderdoni},
  {Colombi}, {Bertin}, {Bouchet}, {Devriendt}, \& {Hatton}}]{momaf}
{Blaizot}, J., {Wadadekar}, Y., {Guiderdoni}, B., {et~al.} 2005, \mnras, 360,
  159

\bibitem[{{Bondi} {et~al.}(2003){Bondi}, {Ciliegi}, {Zamorani}, {Gregorini},
  {Vettolani}, {Parma}, {de Ruiter}, {Le Fevre}, {Arnaboldi}, {Guzzo},
  {Maccagni}, {Scaramella}, {Adami}, {Bardelli}, {Bolzonella}, {Bottini},
  {Cappi}, {Foucaud}, {Franzetti}, {Garilli}, {Gwyn}, {Ilbert}, {Iovino}, {Le
  Brun}, {Marano}, {Marinoni}, {McCracken}, {Meneux}, {Pollo}, {Pozzetti},
  {Radovich}, {Ripepi}, {Rizzo}, {Scodeggio}, {Tresse}, {Zanichelli}, \&
  {Zucca}}]{vvds_radio}
{Bondi}, M., {Ciliegi}, P., {Zamorani}, G., {et~al.} 2003, \aap, 403, 857

\bibitem[{{Bottini} {et~al.}(2005){Bottini}, {Garilli}, {Maccagni}, {Tresse},
  {Le Brun}, {Le F{\`e}vre}, {Picat}, {Scaramella}, {Scodeggio}, {Vettolani},
  {Zanichelli}, {Adami}, {Arnaboldi}, {Arnouts}, {Bardelli}, {Bolzonella},
  {Cappi}, {Charlot}, {Ciliegi}, {Contini}, {Foucaud}, {Franzetti}, {Guzzo},
  {Ilbert}, {Iovino}, {McCracken}, {Marano}, {Marinoni}, {Mathez}, {Mazure},
  {Meneux}, {Merighi}, {Paltani}, {Pollo}, {Pozzetti}, {Radovich}, {Zamorani},
  \& {Zucca}}]{vmmps}
{Bottini}, D., {Garilli}, B., {Maccagni}, D., {et~al.} 2005, \pasp, 117, 996

\bibitem[{{Chiappetti} {et~al.}(2005){Chiappetti}, {Tajer}, {Trinchieri},
  {Maccagni}, {Maraschi}, {Paioro}, {Pierre}, {Surdej}, {Garcet}, {Gosset}, {Le
  F{\`e}vre}, {Bertin}, {McCracken}, {Mellier}, {Foucaud}, {Radovich},
  {Ripepi}, \& {Arnaboldi}}]{vvds_XMDS}
{Chiappetti}, L., {Tajer}, M., {Trinchieri}, G., {et~al.} 2005, \aap, 439, 413

\bibitem[{{Ciliegi} {et~al.}(2005){Ciliegi}, {Zamorani}, {Bondi}, {Pozzetti},
  {Bolzonella}, {Gregorini}, {Garilli}, {Iovino}, {McCracken}, {Mellier},
  {Radovich}, {de Ruiter}, {Parma}, {Bottini}, {Le Brun}, {Le F{\`e}vre},
  {Maccagni}, {Picat}, {Scaramella}, {Scodeggio}, {Tresse}, {Vettolani},
  {Zanichelli}, {Adami}, {Arnaboldi}, {Arnouts}, {Bardelli}, {Cappi},
  {Charlot}, {Contini}, {Foucaud}, {Franzetti}, {Guzzo}, {Ilbert}, {Marano},
  {Marinoni}, {Mathez}, {Mazure}, {Meneux}, {Merighi}, {Merluzzi}, {Paltani},
  {Pollo}, {Zucca}, {Bongiorno}, {Busarello}, {Gavignaud}, {Pell{\`o}},
  {Ripepi}, \& {Rizzo}}]{vvds_radio_2}
{Ciliegi}, P., {Zamorani}, G., {Bondi}, M., {et~al.} 2005, \aap, 441, 879

\bibitem[{{Colless} {et~al.}(2001){Colless}, {Dalton}, {Maddox}, {Sutherland},
  {Norberg}, {Cole}, {Bland-Hawthorn}, {Bridges}, {Cannon}, {Collins}, {Couch},
  {Cross}, {Deeley}, {De Propris}, {Driver}, {Efstathiou}, {Ellis}, {Frenk},
  {Glazebrook}, {Jackson}, {Lahav}, {Lewis}, {Lumsden}, {Madgwick}, {Peacock},
  {Peterson}, {Price}, {Seaborne}, \& {Taylor}}]{2df}
{Colless}, M., {Dalton}, G., {Maddox}, S., {et~al.} 2001, \mnras, 328, 1039

\bibitem[{{Cucciati} {et~al.}(2006){Cucciati}, {Iovino}, {Marinoni}, {Ilbert},
  {Bardelli}, {Franzetti}, {Le F{\`e}vre}, {Pollo}, {Zamorani}, {Cappi},
  {Guzzo}, {McCracken}, {Meneux}, {Scaramella}, {Scodeggio}, {Tresse}, {Zucca},
  {Bottini}, {Garilli}, {Le Brun}, {Maccagni}, {Picat}, {Vettolani},
  {Zanichelli}, {Adami}, {Arnaboldi}, {Arnouts}, {Bolzonella}, {Charlot},
  {Ciliegi}, {Contini}, {Foucaud}, {Gavignaud}, {Marano}, {Mazure}, {Merighi},
  {Paltani}, {Pell{\`o}}, {Pozzetti}, {Radovich}, {Bondi}, {Bongiorno},
  {Busarello}, {de La Torre}, {Gregorini}, {Lamareille}, {Mathez}, {Mellier},
  {Merluzzi}, {Ripepi}, {Rizzo}, {Temporin}, \& {Vergani}}]{vvds_environment}
{Cucciati}, O., {Iovino}, A., {Marinoni}, C., {et~al.} 2006, \aap, 458, 39

\bibitem[{{Daddi} {et~al.}(2004){Daddi}, {Cimatti}, {Renzini}, {Fontana},
  {Mignoli}, {Pozzetti}, {Tozzi}, \& {Zamorani}}]{Daddi}
{Daddi}, E., {Cimatti}, A., {Renzini}, A., {et~al.} 2004, \apj, 617, 746

\bibitem[{{Davis} {et~al.}(2003){Davis}, {Faber}, {Newman}, {Phillips},
  {Ellis}, {Steidel}, {Conselice}, {Coil}, {Finkbeiner}, {Koo}, {Guhathakurta},
  {Weiner}, {Schiavon}, {Willmer}, {Kaiser}, {Luppino}, {Wirth}, {Connolly},
  {Eisenhardt}, {Cooper}, \& {Gerke}}]{deep2}
{Davis}, M., {Faber}, S.~M., {Newman}, J., {et~al.} 2003, in Discoveries and
  Research Prospects from 6- to 10-Meter-Class Telescopes II. Edited by
  Guhathakurta, Puragra. Proceedings of the SPIE, Volume 4834, pp. 161-172
  (2003)., 161--172

\bibitem[{{De Lucia} \& {Blaizot}(2007)}]{delucia}
{De Lucia}, G. \& {Blaizot}, J. 2007, \mnras, 375, 2

\bibitem[{{Fioc} \& {Rocca-Volmerange}(1997)}]{pegase}
{Fioc}, M. \& {Rocca-Volmerange}, B. 1997, \aap, 326, 950

\bibitem[{{Franzetti} {et~al.}(2007){Franzetti}, {Garilli}, {Fumana}, {Paioro},
  {Scodeggio}, S., \& { Scaramella}}]{gossip}
{Franzetti}, P., {Garilli}, B., {Fumana}, M., {et~al.} 2007, in {Proceedings of
  the ESA Workshop "Astronomical Spectroscopy and the Virtual Observatory}, ''

\bibitem[{{Giavalisco} {et~al.}(2004){Giavalisco}, {Ferguson}, {Koekemoer},
  {Dickinson}, {Alexander}, {Bauer}, {Bergeron}, {Biagetti}, {Brandt},
  {Casertano}, {Cesarsky}, {Chatzichristou}, {Conselice}, {Cristiani}, {Da
  Costa}, {Dahlen}, {de Mello}, {Eisenhardt}, {Erben}, {Fall}, {Fassnacht},
  {Fosbury}, {Fruchter}, {Gardner}, {Grogin}, {Hook}, {Hornschemeier}, {Idzi},
  {Jogee}, {Kretchmer}, {Laidler}, {Lee}, {Livio}, {Lucas}, {Madau},
  {Mobasher}, {Moustakas}, {Nonino}, {Padovani}, {Papovich}, {Park},
  {Ravindranath}, {Renzini}, {Richardson}, {Riess}, {Rosati}, {Schirmer},
  {Schreier}, {Somerville}, {Spinrad}, {Stern}, {Stiavelli}, {Strolger},
  {Urry}, {Vandame}, {Williams}, \& {Wolf}}]{goods}
{Giavalisco}, M., {Ferguson}, H.~C., {Koekemoer}, A.~M., {et~al.} 2004, \apjl,
  600, L93

\bibitem[{{Guzzo} {et~al.}(2008){Guzzo}, {Pierleoni}, {Meneux}, {Branchini},
  {Le F{\`e}vre}, {Marinoni}, {Garilli}, {Blaizot}, {De Lucia}, {Pollo},
  {McCracken}, {Bottini}, {Le Brun}, {Maccagni}, {Picat}, {Scaramella},
  {Scodeggio}, {Tresse}, {Vettolani}, {Zanichelli}, {Adami}, {Arnouts},
  {Bardelli}, {Bolzonella}, {Bongiorno}, {Cappi}, {Charlot}, {Ciliegi},
  {Contini}, {Cucciati}, {de La Torre}, {Dolag}, {Foucaud}, {Franzetti},
  {Gavignaud}, {Ilbert}, {Iovino}, {Lamareille}, {Marano}, {Mazure}, {Memeo},
  {Merighi}, {Moscardini}, {Paltani}, {Pell{\`o}}, {Perez-Montero}, {Pozzetti},
  {Radovich}, {Vergani}, {Zamorani}, \& {Zucca}}]{guzzo_nature}
{Guzzo}, L., {Pierleoni}, M., {Meneux}, B., {et~al.} 2008, \nat, 451, 541

\bibitem[{{Ilbert} {et~al.}(2006){Ilbert}, {Cucciati}, {Marinoni}, {Tresse},
  {Le Fevre}, {Zamorani}, {Bardelli}, {Iovino}, {Zucca}, {Arnouts}, {Bottini},
  {Garilli}, {Le Brun}, {Maccagni}, {Picat}, {Scaramella}, {Scodeggio},
  {Vettolani}, {Zanichelli}, {Adami}, {Bolzonella}, {Cappi}, {Charlot},
  {Ciliegi}, {Contini}, {Foucaud}, {Franzetti}, {Gavignaud}, {Guzzo}, {Marano},
  {Mazure}, {McCracken}, {Meneux}, {Merighi}, {Paltani}, {Pello}, {Pollo},
  {Pozzetti}, {Radovich}, {Bondi}, {Bongiorno}, {Busarello}, {De La Torre},
  {Gregorini}, {Lamareille}, {Mathez}, {Mellier}, {Merluzzi}, {Ripepi},
  {Rizzo}, \& {Vergani}}]{vvds_lf_environment}
{Ilbert}, O., {Cucciati}, O., {Marinoni}, C., {et~al.} 2006, ArXiv Astrophysics
  e-prints

\bibitem[{{Ilbert} {et~al.}(2005){Ilbert}, {Tresse}, {Zucca}, {Bardelli},
  {Arnouts}, {Zamorani}, {Pozzetti}, {Bottini}, {Garilli}, {Le Brun}, {Le
  F{\`e}vre}, {Maccagni}, {Picat}, {Scaramella}, {Scodeggio}, {Vettolani},
  {Zanichelli}, {Adami}, {Arnaboldi}, {Bolzonella}, {Cappi}, {Charlot},
  {Contini}, {Foucaud}, {Franzetti}, {Gavignaud}, {Guzzo}, {Iovino},
  {McCracken}, {Marano}, {Marinoni}, {Mathez}, {Mazure}, {Meneux}, {Merighi},
  {Paltani}, {Pello}, {Pollo}, {Radovich}, {Bondi}, {Bongiorno}, {Busarello},
  {Ciliegi}, {Lamareille}, {Mellier}, {Merluzzi}, {Ripepi}, \&
  {Rizzo}}]{vvds_lf}
{Ilbert}, O., {Tresse}, L., {Zucca}, E., {et~al.} 2005, \aap, 439, 863

\bibitem[{{Iovino} {et~al.}(2005){Iovino}, {McCracken}, {Garilli}, {Foucaud},
  {Le F{\`e}vre}, {Maccagni}, {Saracco}, {Bardelli}, {Busarello}, {Scodeggio},
  {Zanichelli}, {Paioro}, {Bottini}, {Le Brun}, {Picat}, {Scaramella},
  {Tresse}, {Vettolani}, {Adami}, {Arnaboldi}, {Arnouts}, {Bolzonella},
  {Cappi}, {Charlot}, {Ciliegi}, {Contini}, {Franzetti}, {Gavignaud}, {Guzzo},
  {Ilbert}, {Marano}, {Marinoni}, {Mazure}, {Meneux}, {Merighi}, {Paltani},
  {Pell{\`o}}, {Pollo}, {Pozzetti}, {Radovich}, {Zamorani}, {Zucca}, {Bertin},
  {Bondi}, {Bongiorno}, {Cucciati}, {Gregorini}, {Mathez}, {Mellier},
  {Merluzzi}, {Ripepi}, \& {Rizzo}}]{vvds_imaging_jk}
{Iovino}, A., {McCracken}, H.~J., {Garilli}, B., {et~al.} 2005, \aap, 442, 423

\bibitem[{{Koo}(1995)}]{deep}
{Koo}, D. 1995, in Wide Field Spectroscopy and the Distant Universe, ed. S.~J.
  {Maddox} \& A.~{Aragon-Salamanca}, 55--+

\bibitem[{{Le F{\`e}vre} {et~al.}(2005{\natexlab{a}}){Le F{\`e}vre}, {Guzzo},
  {Meneux}, {Pollo}, {Cappi}, {Colombi}, {Iovino}, {Marinoni}, {McCracken},
  {Scaramella}, {Bottini}, {Garilli}, {Le Brun}, {Maccagni}, {Picat},
  {Scodeggio}, {Tresse}, {Vettolani}, {Zanichelli}, {Adami}, {Arnaboldi},
  {Arnouts}, {Bardelli}, {Blaizot}, {Bolzonella}, {Charlot}, {Ciliegi},
  {Contini}, {Foucaud}, {Franzetti}, {Gavignaud}, {Ilbert}, {Marano}, {Mathez},
  {Mazure}, {Merighi}, {Paltani}, {Pell{\`o}}, {Pozzetti}, {Radovich},
  {Zamorani}, {Zucca}, {Bondi}, {Bongiorno}, {Busarello}, {Lamareille},
  {Mellier}, {Merluzzi}, {Ripepi}, \& {Rizzo}}]{vvds_clus}
{Le F{\`e}vre}, O., {Guzzo}, L., {Meneux}, B., {et~al.} 2005{\natexlab{a}},
  \aap, 439, 877

\bibitem[{{Le Fevre} {et~al.}(1996){Le Fevre}, {Hudon}, {Lilly}, {Crampton},
  {Hammer}, \& {Tresse}}]{cfrs_clustering}
{Le Fevre}, O., {Hudon}, D., {Lilly}, S.~J., {et~al.} 1996, \apj, 461, 534

\bibitem[{{Le F{\`e}vre} {et~al.}(2005{\natexlab{b}}){Le F{\`e}vre}, {Paltani},
  {Arnouts}, {Charlot}, {Foucaud}, {Ilbert}, {McCracken}, {Zamorani},
  {Bottini}, {Garilli}, {Le Brun}, {Maccagni}, {Picat}, {Scaramella},
  {Scodeggio}, {Tresse}, {Vettolani}, {Zanichelli}, {Adami}, {Bardelli},
  {Bolzonella}, {Cappi}, {Ciliegi}, {Contini}, {Franzetti}, {Gavignaud},
  {Guzzo}, {Iovino}, {Marano}, {Marinoni}, {Mazure}, {Meneux}, {Merighi},
  {Pell{\`o}}, {Pollo}, {Pozzetti}, {Radovich}, {Zucca}, {Arnaboldi}, {Bondi},
  {Bongiorno}, {Busarello}, {Gregorini}, {Lamareille}, {Mathez}, {Mellier},
  {Merluzzi}, {Ripepi}, \& {Rizzo}}]{vvds_nature}
{Le F{\`e}vre}, O., {Paltani}, S., {Arnouts}, S., {et~al.} 2005{\natexlab{b}},
  \nat, 437, 519

\bibitem[{{Le F{\`e}vre} {et~al.}(2005{\natexlab{c}}){Le F{\`e}vre},
  {Vettolani}, {Garilli}, {Tresse}, {Bottini}, {Le Brun}, {Maccagni}, {Picat},
  {Scaramella}, {Scodeggio}, {Zanichelli}, {Adami}, {Arnaboldi}, {Arnouts},
  {Bardelli}, {Bolzonella}, {Cappi}, {Charlot}, {Ciliegi}, {Contini},
  {Foucaud}, {Franzetti}, {Gavignaud}, {Guzzo}, {Ilbert}, {Iovino},
  {McCracken}, {Marano}, {Marinoni}, {Mathez}, {Mazure}, {Meneux}, {Merighi},
  {Paltani}, {Pell{\`o}}, {Pollo}, {Pozzetti}, {Radovich}, {Zamorani}, {Zucca},
  {Bondi}, {Bongiorno}, {Busarello}, {Lamareille}, {Mellier}, {Merluzzi},
  {Ripepi}, \& {Rizzo}}]{vvds_main}
{Le F{\`e}vre}, O., {Vettolani}, G., {Garilli}, B., {et~al.}
  2005{\natexlab{c}}, \aap, 439, 845

\bibitem[{{LeFevre} {et~al.}(2003){LeFevre}, {Saisse}, {Mancini}, {Brau-Nogue},
  {Caputi}, {Castinel}, {D'Odorico}, {Garilli}, {Kissler-Patig}, {Lucuix},
  {Mancini}, {Pauget}, {Sciarretta}, {Scodeggio}, {Tresse}, \&
  {Vettolani}}]{vvds_tech}
{LeFevre}, O., {Saisse}, M., {Mancini}, D., {et~al.} 2003, in Proceedings of
  the SPIE, ed. M.~{Iye} \& A.~F.~M. {Moorwood}, Vol. 4841, 1670--1681

\bibitem[{{Li} {et~al.}(2006){Li}, {Kauffmann}, {Jing}, {White}, {B{\"o}rner},
  \& {Cheng}}]{sdss_clus2}
{Li}, C., {Kauffmann}, G., {Jing}, Y.~P., {et~al.} 2006, \mnras, 368, 21

\bibitem[{{Lonsdale} {et~al.}(2003){Lonsdale}, {Smith}, {Rowan-Robinson},
  {Surace}, {Shupe}, {Xu}, {Oliver}, {Padgett}, {Fang}, {Conrow},
  {Franceschini}, {Gautier}, {Griffin}, {Hacking}, {Masci}, {Morrison},
  {O'Linger}, {Owen}, {P{\'e}rez-Fournon}, {Pierre}, {Puetter}, {Stacey},
  {Castro}, {Polletta}, {Farrah}, {Jarrett}, {Frayer}, {Siana}, {Babbedge},
  {Dye}, {Fox}, {Gonzalez-Solares}, {Salaman}, {Berta}, {Condon}, {Dole}, \&
  {Serjeant}}]{vvds_swire}
{Lonsdale}, C.~J., {Smith}, H.~E., {Rowan-Robinson}, M., {et~al.} 2003, \pasp,
  115, 897

\bibitem[{{Madgwick} {et~al.}(2003){Madgwick}, {Hawkins}, {Lahav}, {Maddox},
  {Norberg}, {Peacock}, {Baldry}, {Baugh}, {Bland-Hawthorn}, {Bridges},
  {Cannon}, {Cole}, {Colless}, {Collins}, {Couch}, {Dalton}, {De Propris},
  {Driver}, {Efstathiou}, {Ellis}, {Frenk}, {Glazebrook}, {Jackson}, {Lewis},
  {Lumsden}, {Peterson}, {Sutherland}, \& {Taylor}}]{2df_clus_type}
{Madgwick}, D.~S., {Hawkins}, E., {Lahav}, O., {et~al.} 2003, \mnras, 344, 847

\bibitem[{{Marinoni} {et~al.}(2005){Marinoni}, {Le F{\`e}vre}, {Meneux},
  {Iovino}, {Pollo}, {Ilbert}, {Zamorani}, {Guzzo}, {Mazure}, {Scaramella},
  {Cappi}, {McCracken}, {Bottini}, {Garilli}, {Le Brun}, {Maccagni}, {Picat},
  {Scodeggio}, {Tresse}, {Vettolani}, {Zanichelli}, {Adami}, {Arnouts},
  {Bardelli}, {Blaizot}, {Bolzonella}, {Charlot}, {Ciliegi}, {Contini},
  {Foucaud}, {Franzetti}, {Gavignaud}, {Marano}, {Mathez}, {Merighi},
  {Paltani}, {Pell{\`o}}, {Pozzetti}, {Radovich}, {Zucca}, {Bondi},
  {Bongiorno}, {Busarello}, {Colombi}, {Cucciati}, {Lamareille}, {Mellier},
  {Merluzzi}, {Ripepi}, \& {Rizzo}}]{vvds_bias}
{Marinoni}, C., {Le F{\`e}vre}, O., {Meneux}, B., {et~al.} 2005, \aap, 442, 801

\bibitem[{{McCracken} {et~al.}(2003){McCracken}, {Radovich}, {Bertin},
  {Mellier}, {Dantel-Fort}, {Le F{\` e}vre}, {Cuillandre}, {Gwyn}, {Foucaud},
  \& {Zamorani}}]{vvds_imaging_f02}
{McCracken}, H.~J., {Radovich}, M., {Bertin}, E., {et~al.} 2003, \aap, 410, 17

\bibitem[{{Meneux} {et~al.}(2008){Meneux}, {Guzzo}, {Garilli}, {Le F{\`e}vre},
  {Pollo}, {Blaizot}, {De Lucia}, {Bolzonella}, {Lamareille}, {Pozzetti},
  {Cappi}, {Iovino}, {Marinoni}, {McCracken}, {de La Torre}, {Bottini}, {Le
  Brun}, {Maccagni}, {Picat}, {Scaramella}, {Scodeggio}, {Tresse}, {Vettolani},
  {Zanichelli}, {Abbas}, {Adami}, {Arnouts}, {Bardelli}, {Bongiorno},
  {Charlot}, {Ciliegi}, {Contini}, {Cucciati}, {Foucaud}, {Franzetti},
  {Gavignaud}, {Ilbert}, {Marano}, {Mazure}, {Merighi}, {Paltani}, {Pell{\`o}},
  {Radovich}, {Vergani}, {Zamorani}, \& {Zucca}}]{vvds_clus_mass}
{Meneux}, B., {Guzzo}, L., {Garilli}, B., {et~al.} 2008, \aap, 478, 299

\bibitem[{{Meneux} {et~al.}(2006){Meneux}, {Le F{\`e}vre}, {Guzzo}, {Pollo},
  {Cappi}, {Ilbert}, {Iovino}, {Marinoni}, {McCracken}, {Bottini}, {Garilli},
  {Le Brun}, {Maccagni}, {Picat}, {Scaramella}, {Scodeggio}, {Tresse},
  {Vettolani}, {Zanichelli}, {Adami}, {Arnouts}, {Arnaboldi}, {Bardelli},
  {Bolzonella}, {Charlot}, {Ciliegi}, {Contini}, {Foucaud}, {Franzetti},
  {Gavignaud}, {Marano}, {Mazure}, {Merighi}, {Paltani}, {Pell{\`o}},
  {Pozzetti}, {Radovich}, {Zamorani}, {Zucca}, {Bondi}, {Bongiorno},
  {Busarello}, {Cucciati}, {Gregorini}, {Lamareille}, {Mathez}, {Mellier},
  {Merluzzi}, {Ripepi}, \& {Rizzo}}]{vvds_clus_type}
{Meneux}, B., {Le F{\`e}vre}, O., {Guzzo}, L., {et~al.} 2006, \aap, 452, 387

\bibitem[{{Norberg} {et~al.}(2002){Norberg}, {Baugh}, {Hawkins}, {Maddox},
  {Madgwick}, {Lahav}, {Cole}, {Frenk}, {Baldry}, {Bland-Hawthorn}, {Bridges},
  {Cannon}, {Colless}, {Collins}, {Couch}, {Dalton}, {De Propris}, {Driver},
  {Efstathiou}, {Ellis}, {Glazebrook}, {Jackson}, {Lewis}, {Lumsden},
  {Peacock}, {Peterson}, {Sutherland}, \& {Taylor}}]{2df_clus2}
{Norberg}, P., {Baugh}, C.~M., {Hawkins}, E., {et~al.} 2002, \mnras, 332, 827

\bibitem[{{Norberg} {et~al.}(2001){Norberg}, {Baugh}, {Hawkins}, {Maddox},
  {Peacock}, {Cole}, {Frenk}, {Bland-Hawthorn}, {Bridges}, {Cannon}, {Colless},
  {Collins}, {Couch}, {Dalton}, {De Propris}, {Driver}, {Efstathiou}, {Ellis},
  {Glazebrook}, {Jackson}, {Lahav}, {Lewis}, {Lumsden}, {Madgwick}, {Peterson},
  {Sutherland}, \& {Taylor}}]{2df_clus1}
{Norberg}, P., {Baugh}, C.~M., {Hawkins}, E., {et~al.} 2001, \mnras, 328, 64

\bibitem[{{Peebles}(1980)}]{peebles}
{Peebles}, P.~J.~E. 1980, {The large-scale structure of the universe}
  (Princeton University Press)

\bibitem[{{Pierre} {et~al.}(2004){Pierre}, {Valtchanov}, {Altieri}, {Andreon},
  {Bolzonella}, {Bremer}, {Disseau}, {Dos Santos}, {Gandhi}, {Jean}, {Pacaud},
  {Read}, {Refregier}, {Willis}, {Adami}, {Alloin}, {Birkinshaw}, {Chiappetti},
  {Cohen}, {Detal}, {Duc}, {Gosset}, {Hjorth}, {Jones}, {LeFevre}, {Lonsdale},
  {Maccagni}, {Mazure}, {McBreen}, {McCracken}, {Mellier}, {Ponman},
  {Quintana}, {Rottgering}, {Smette}, {Surdej}, {Starck}, {Vigroux}, \&
  {White}}]{vvds_LSS}
{Pierre}, M., {Valtchanov}, I., {Altieri}, B., {et~al.} 2004, Journal of
  Cosmology and Astro-Particle Physics, 9, 11

\bibitem[{{Pollo} {et~al.}(2006){Pollo}, {Guzzo}, {Le F{\`e}vre}, {Meneux},
  {Cappi}, {Franzetti}, {Iovino}, {McCracken}, {Marinoni}, {Zamorani},
  {Bottini}, {Garilli}, {Le Brun}, {Maccagni}, {Picat}, {Scaramella},
  {Scodeggio}, {Tresse}, {Vettolani}, {Zanichelli}, {Adami}, {Arnouts},
  {Bardelli}, {Bolzonella}, {Charlot}, {Ciliegi}, {Contini}, {Foucaud},
  {Gavignaud}, {Ilbert}, {Marano}, {Mazure}, {Merighi}, {Paltani}, {Pell{\`o}},
  {Pozzetti}, {Radovich}, {Zucca}, {Bondi}, {Bongiorno}, {Busarello},
  {Cucciati}, {Gregorini}, {Lamareille}, {Mathez}, {Mellier}, {Merluzzi},
  {Ripepi}, \& {Rizzo}}]{vvds_clus_lum}
{Pollo}, A., {Guzzo}, L., {Le F{\`e}vre}, O., {et~al.} 2006, \aap, 451, 409

\bibitem[{{Pozzetti} {et~al.}(2007){Pozzetti}, {Bolzonella}, {Lamareille},
  {Zamorani}, {Franzetti}, {Le F{\`e}vre}, {Iovino}, {Temporin}, {Ilbert},
  {Arnouts}, {Charlot}, {Brinchmann}, {Zucca}, {Tresse}, {Scodeggio}, {Guzzo},
  {Bottini}, {Garilli}, {Le Brun}, {Maccagni}, {Picat}, {Scaramella},
  {Vettolani}, {Zanichelli}, {Adami}, {Bardelli}, {Cappi}, {Ciliegi},
  {Contini}, {Foucaud}, {Gavignaud}, {McCracken}, {Marano}, {Marinoni},
  {Mazure}, {Meneux}, {Merighi}, {Paltani}, {Pell{\`o}}, {Pollo}, {Radovich},
  {Bondi}, {Bongiorno}, {Cucciati}, {de La Torre}, {Gregorini}, {Mellier},
  {Merluzzi}, {Vergani}, \& {Walcher}}]{vvds_massfunc}
{Pozzetti}, L., {Bolzonella}, M., {Lamareille}, F., {et~al.} 2007, \aap, 474,
  443

\bibitem[{{Radovich} {et~al.}(2004){Radovich}, {Arnaboldi}, {Ripepi},
  {Massarotti}, {McCracken}, {Mellier}, {Bertin}, {Zamorani}, {Adami},
  {Bardelli}, {Le F{\` e}vre}, {Foucaud}, {Garilli}, {Scaramella}, {Vettolani},
  {Zanichelli}, \& {Zucca}}]{vvds_imaging_u}
{Radovich}, M., {Arnaboldi}, M., {Ripepi}, V., {et~al.} 2004, \aap, 417, 51

\bibitem[{{Schiminovich} {et~al.}(2005){Schiminovich}, {Ilbert}, {Arnouts},
  {Milliard}, {Tresse}, {Le F{\`e}vre}, {Treyer}, {Wyder}, {Budav{\'a}ri},
  {Zucca}, {Zamorani}, {Martin}, {Adami}, {Arnaboldi}, {Bardelli}, {Barlow},
  {Bianchi}, {Bolzonella}, {Bottini}, {Byun}, {Cappi}, {Contini}, {Charlot},
  {Donas}, {Forster}, {Foucaud}, {Franzetti}, {Friedman}, {Garilli},
  {Gavignaud}, {Guzzo}, {Heckman}, {Hoopes}, {Iovino}, {Jelinsky}, {Le Brun},
  {Lee}, {Maccagni}, {Madore}, {Malina}, {Marano}, {Marinoni}, {McCracken},
  {Mazure}, {Meneux}, {Morrissey}, {Neff}, {Paltani}, {Pell{\`o}}, {Picat},
  {Pollo}, {Pozzetti}, {Radovich}, {Rich}, {Scaramella}, {Scodeggio},
  {Seibert}, {Siegmund}, {Small}, {Szalay}, {Vettolani}, {Welsh}, {Xu}, \&
  {Zanichelli}}]{vvds_galex_lumden}
{Schiminovich}, D., {Ilbert}, O., {Arnouts}, S., {et~al.} 2005, \apjl, 619, L47

\bibitem[{{Scodeggio} {et~al.}(2005){Scodeggio}, {Franzetti}, {Garilli},
  {Zanichelli}, {Paltani}, {Maccagni}, {Bottini}, {Le Brun}, {Contini},
  {Scaramella}, {Adami}, {Bardelli}, {Zucca}, {Tresse}, {Ilbert}, {Foucaud},
  {Iovino}, {Merighi}, {Zamorani}, {Gavignaud}, {Rizzo}, {McCracken}, {Le
  F{\`e}vre}, {Picat}, {Vettolani}, {Arnaboldi}, {Arnouts}, {Bolzonella},
  {Cappi}, {Charlot}, {Ciliegi}, {Guzzo}, {Marano}, {Marinoni}, {Mathez},
  {Mazure}, {Meneux}, {Pell{\`o}}, {Pollo}, {Pozzetti}, \& {Radovich}}]{vipgi}
{Scodeggio}, M., {Franzetti}, P., {Garilli}, B., {et~al.} 2005, \pasp, 117,
  1284

\bibitem[{{Shepherd} {et~al.}(2001){Shepherd}, {Carlberg}, {Yee}, {Morris},
  {Lin}, {Sawicki}, {Hall}, \& {Patton}}]{cnoc_clustering}
{Shepherd}, C.~W., {Carlberg}, R.~G., {Yee}, H.~K.~C., {et~al.} 2001, \apj,
  560, 72

\bibitem[{{Somerville} {et~al.}(2004){Somerville}, {Lee}, {Ferguson},
  {Gardner}, {Moustakas}, \& {Giavalisco}}]{somerville}
{Somerville}, R.~S., {Lee}, K., {Ferguson}, H.~C., {et~al.} 2004, \apjl, 600,
  L171

\bibitem[{{Springel} {et~al.}(2005){Springel}, {White}, {Jenkins}, {Frenk},
  {Yoshida}, {Gao}, {Navarro}, {Thacker}, {Croton}, {Helly}, {Peacock}, {Cole},
  {Thomas}, {Couchman}, {Evrard}, {Colberg}, \& {Pearce}}]{millennium}
{Springel}, V., {White}, S.~D.~M., {Jenkins}, A., {et~al.} 2005, \nat, 435, 629

\bibitem[{{Steidel} {et~al.}(1998){Steidel}, {Adelberger}, {Dickinson},
  {Giavalisco}, {Pettini}, \& {Kellogg}}]{steidel98}
{Steidel}, C.~C., {Adelberger}, K.~L., {Dickinson}, M., {et~al.} 1998, \apj,
  492, 428

\bibitem[{{Warren} {et~al.}(2007){Warren}, {Hambly}, {Dye}, {Almaini}, {Cross},
  {Edge}, {Foucaud}, {Hewett}, {Hodgkin}, {Irwin}, {Jameson}, {Lawrence},
  {Lucas}, {Adamson}, {Bandyopadhyay}, {Bryant}, {Collins}, {Davis}, {Dunlop},
  {Emerson}, {Evans}, {Gonzales-Solares}, {Hirst}, {Jarvis}, {Kendall}, {Kerr},
  {Leggett}, {Lewis}, {Mann}, {McLure}, {McMahon}, {Mortlock}, {Rawlings},
  {Read}, {Riello}, {Simpson}, {Smith}, {Sutorius}, {Targett}, \&
  {Varricatt}}]{UKIDSS}
{Warren}, S.~J., {Hambly}, N.~C., {Dye}, S., {et~al.} 2007, \mnras, 375, 213

\bibitem[{{White} \& {Rees}(1978)}]{white_rees}
{White}, S.~D.~M. \& {Rees}, M.~J. 1978, \mnras, 183, 341

\bibitem[{{Zehavi} {et~al.}(2005){Zehavi}, {Zheng}, {Weinberg}, {Frieman},
  {Berlind}, {Blanton}, {Scoccimarro}, {Sheth}, {Strauss}, {Kayo}, {Suto},
  {Fukugita}, {Nakamura}, {Bahcall}, {Brinkmann}, {Gunn}, {Hennessy},
  {Ivezi{\'c}}, {Knapp}, {Loveday}, {Meiksin}, {Schlegel}, {Schneider},
  {Szapudi}, {Tegmark}, {Vogeley}, \& {York}}]{sdss_clus1}
{Zehavi}, I., {Zheng}, Z., {Weinberg}, D.~H., {et~al.} 2005, \apj, 630, 1

\bibitem[{{Zucca} {et~al.}(2006){Zucca}, {Ilbert}, {Bardelli}, {Tresse},
  {Zamorani}, {Arnouts}, {Pozzetti}, {Bolzonella}, {McCracken}, {Bottini},
  {Garilli}, {Le Brun}, {Le F{\`e}vre}, {Maccagni}, {Picat}, {Scaramella},
  {Scodeggio}, {Vettolani}, {Zanichelli}, {Adami}, {Arnaboldi}, {Cappi},
  {Charlot}, {Ciliegi}, {Contini}, {Foucaud}, {Franzetti}, {Gavignaud},
  {Guzzo}, {Iovino}, {Marano}, {Marinoni}, {Mazure}, {Meneux}, {Merighi},
  {Paltani}, {Pell{\`o}}, {Pollo}, {Radovich}, {Bondi}, {Bongiorno},
  {Busarello}, {Cucciati}, {Gregorini}, {Lamareille}, {Mathez}, {Mellier},
  {Merluzzi}, {Ripepi}, \& {Rizzo}}]{vvds_lf_type}
{Zucca}, E., {Ilbert}, O., {Bardelli}, S., {et~al.} 2006, \aap, 455, 879

\end{thebibliography}
~\\
\end{document}